\documentclass{aa}  

\usepackage{graphicx}
\usepackage{txfonts}

\graphicspath{ {./figures/} }
\usepackage{grffile}
\usepackage[hidelinks,colorlinks=true,linkcolor=blue,anchorcolor=black,citecolor=blue,filecolor=black,menucolor=black,runcolor=black,urlcolor=black]{hyperref}
\begin{document} 

    \title{Ices in planet-forming disks: Self-consistent ice opacities in disk models}


   \author{Aditya M. Arabhavi \inst{1,2,3,6} \and
            Peter Woitke \inst{2,6} \and
            St\'ephanie M. Cazaux \inst{1,5} \and
            Inga Kamp \inst{3} \and
            Christian Rab \inst{4,7} \and
            Wing-Fai Thi \inst{4}
          }

   \institute{Faculty of Aerospace Engineering, Delft University of Technology, Delft, The Netherlands
              \and
              School of Physics \& Astronomy, University of St. Andrews, North Haugh, St. Andrews KY16 9SS, UK
              \and
              Kapteyn Astronomical Institute, University of Groningen, P.O. Box 800, 9700 AV Groningen, The Netherlands;\\ \email{\href{mailto:arabhavi@astro.rug.nl}{arabhavi@astro.rug.nl}}
              \and
              Max Planck Institute for Extraterrestrial Physics, Giessenbachstrasse 1, 85741 Garching, Germany
              \and
              Leiden Observatory, Leiden University, P.O. Box 9513, NL 2300 RA Leiden, The Netherlands
              \and 
              Centre for Exoplanet Science, University of St Andrews, North Haugh, St Andrews, KY16 9SS, UK
              \and
              Universit\"ats-Sternwarte, Fakult\"at f\"ur Physik,   Ludwig-Maximilians-Universit\"at M\"unchen, Scheinerstr.~1, 81679 M\"unchen, Germany
             }

   \date{Received July 21, 2021; accepted August 18, 2022}
    \titlerunning{Ices in planet-forming disks}
    \authorrunning{A. M. Arabhavi et al.}
 
  \abstract
   {In cold and shielded environments, molecules freeze out on dust grain surfaces to form ices such as {$\rm H_2O$}, {$\rm CO$}, {$\rm CO_2$}, {$\rm CH_4$}, {$\rm CH_3OH$}, and {$\rm NH_3$}. In protoplanetary disks, such conditions are present in the midplane regions beyond the snowline, but the exact radial and vertical extension depend on disk mass, geometry, and stellar ultra-violet irradiation.}
   {The goal of this work is to present a computationally efficient method to compute ice and bare-grain opacities in protoplanetary disk models consistently with the chemistry and to investigate the effect of ice opacities on the physico-chemical state and optical appearance of the disk.}
   {A matrix of Mie efficiencies is pre-calculated for different ice species and thicknesses, from which the position dependent opacities of icy grains are then interpolated. This is implemented in the {P{\tiny RO}D{\tiny I}M{\tiny O}} code by a self-consistent solution of ice opacities and the local composition of ices, which are obtained from our chemical network.}
   {Locally, the opacity can change significantly, for example, an increase by a factor of more than 200 in the midplane, especially at ultra-violet and optical wavelengths, due to ice formation. This is mainly due to changes in the size distribution of dust grains resulting from ice formation. However, since the opacity only changes in the optically thick regions of the disk, the thermal disk structure does not change significantly. For the same reason, the spectral energy distributions (SEDs) computed with our disk models with ice opacities generally show only faint ice emission features at far-IR wavelengths. The ice absorption features are only seen in the edge-on orientation. The assumption made on how the ice is distributed across the grain size distribution (ice power law) influences the far-infrared and millimeter slope of the SED. The ice features and their strengths are influenced by the ice power law and the type of chemistry. Our models predict stronger ice features for observations that can spatially resolve the disk, particularly in absorption.}
   {}


   \maketitle
%

\section{Introduction}
\label{sec:intro}
Protoplanetary disks are made up of gas and dust with different spatial distribution and temperature structures. The midplane is generally cold and dense for both dust and gas components \citep[e.g.][]{2007prpl.conf..751B,2021PhR...893....1O}. In those cold regions, which are well shielded from both stellar and interstellar ultra-violet (UV) radiation fields, gas phase species can accrete onto dust grain surfaces to form ices. The local thermo-chemical conditions in these regions determine the volume and chemical composition of the forming ice. This ice accretion affects the grain size and composition which are important for opacity calculations \citep{ossenkopf1994dust,allamandola1999evolution,boogert2004interstellar,oberg2009complex}.

Observational data of dust in the diffuse interstellar medium, molecular clouds, and young stellar objects show the presence of ice covering the dust grains \citep{merrill1976infrared,whittet1983interstellar,whittet1998detection,chiar2000composition,chiar2002hydrocarbons,boogert2008c2d,goto2018first}. For protoplanetary disks, the resolution, spectral range, and sensitivity requirements limit the observational capabilities. Yet some observations of {$\rm H_2O$} ice in absorption at 3$\mu$m \citep[for example][]{terada2007detection,2009ApJ...690L.110H,honda2016water,terada2017multi} and emission at 47\,$\mu$m and 63\,$\mu$m \citep{mcclure2012probing,mcclure2015detections,min2016abundance} are available. Thus, disk modeling becomes important for understanding the properties of icy dust in disks as well as to design observing strategies for observing ices. 

Observations, such as \citet{terada2007detection} and \citet{2012ApJ...753...19T}, indicate that the observed profile of an ice feature (for example the 3~$\mu$m {$\rm H_2O$} ice feature) is a combination of diverse factors. In addition to the contributions of the ice species themselves, which include presence of different ice species such as {$\rm CO$}, {$\rm CH_4$}, {$\rm NH_3$}, {$\rm CH_3OH$} \citep{brooke1999new,2001A&A...365..144D,dartois2002combined}, optical behavior of large grains \citep{smith1989absorption,boogert2015observations} and thermal state of the ices \citep{boogert2015observations,sandford1988condensation,hudgins1993mid,palumbo19932140} contribute to the shape of the ice feature. Grain growth is also expected in cold and dense environments \citep[e.g.][]{2012ApJ...753...19T,terada2012adaptive,mcclure2012probing}.

Generally, disk models \citep[e.g.][]{nomura2005molecular,d2006effects,semenov2011chemical,2012A&A...541A..91B,woitke2016consistent} assume a grain size distribution with a minimum and maximum grain size, and a power law with index $a_{\rm pow}$ that is either obtained by fitting the observations or set to 3.5 \citep[MRN-distribution,][]{mathis1977size}. These grains are considered to be composed of refractory materials such as silicates and conducting materials. Some models consider the presence of ices by using ice opacities \citep[e.g.][]{nomura2005molecular,pontoppidan2005ices,mcclure2012probing,2018A&A...617A...1K}. However, some of these models \citep{nomura2005molecular,d2006effects,mcclure2012probing,mcclure2015detections} consider segregated grain distribution, that is, having separate size distribution of ice grains apart from refractory grains. Moreover, for calculating the opacities, these models \citep{nomura2005molecular,d2006effects,mcclure2015detections} either assume that ice exists everywhere below sublimation temperature, neglecting photodesorption, or do not consistently solve for photodesorption and ice abundance. Further, \citet{nomura2005molecular}, \citet{d2006effects}, \citet{mcclure2012probing}, \citet{mcclure2015detections}, and \citet{2018A&A...617A...1K} consider only {$\rm H_2O$} ice and \citet{pontoppidan2005ices} consider {$\rm H_2O$}, {$\rm CO$}, and {$\rm CO_2$} ice species for opacity calculations. Segregated grain distribution do not account for increase in grain size due to ice accretion. Considering only $\rm H_2O$ ice underestimates the effect of ice accretion on size distribution of grains. A more recent study of ice opacities in disk models is presented in \citet{B21}, where six ice species ($\rm H_2O$, $\rm CO$, $\rm CO_2$, $\rm CH_3OH$, $\rm NH_3$, and $\rm CH_4$) are considered along with two populations of refractory grains. The disk region is spatially classified into 25 zones which can be approximated with the same opacities. Thus avoiding the computationally intensive, yet more accurate, point-by-point opacity calculation. The effects of an interstellar UV field is also neglected. The model produces substantial ice absorption features including the 3~$\mu$m water ice band in disks that are not edge on. A more consistent treatment of the effect of ice on dust composition and size, hence, will help improve the understanding of ice observations in disks.

\citet{ossenkopf1994dust} show, for a fixed volume of water ice, that the opacity changes when ice mantles are present on dust grains, which is not taken into consideration by many disk models. \citet{2018A&A...617A...1K} assumes that {$\rm H_2O$} ice is present in region defined by the snowline \citep{min2016abundance} which takes into account the temperature, pressure, and UV radiation field. The ice is distributed as ice mantles on top of underlying bare grain size distribution. Ices are assumed to be a constant mass fraction of the total dust mass, thus the ice abundance is not informed by the local chemical state. This leads to having too much ice in the upper midplane or too little ice in the inner midplane. \citet{pontoppidan2005ices} use a single refractory grain size population, upon which ice mantle composed of {$\rm H_2O$}, {$\rm CO$}, and {$\rm CO_2$} ices is present. However, the model considers that ices evaporate at 90K (20K for CO) and also the local ice abundances are obtained by ice mixtures of fixed proportions ({$\rm H_2O$}:{$\rm CO_2$}:{$\rm CO$}={$\rm 1$}:{$\rm 0.2$}:{$\rm 0.03$} and inclusions of {$\rm CO_2$}:{$\rm CO$}={$\rm 1$}:{$\rm 0.7$}). In this paper, we present a model in which, for the first time, the ice chemistry, opacities, and radiative transfer are treated self-consistently.

We present here a computationally efficient method of including position dependent opacities in disk models, to account for ice abundances based on local thermo-chemical state, and discuss the implications for observations. In Sect. \ref{sec:prodimo} the description of the disk model, {P{\small RO}D{\small I}M{\small O}}, is provided in which the position dependent ice opacities are implemented. In Sect.~\ref{sec:opacity}, the method of calculating position dependent opacities is explained along with the verification of the calculation. The effect of the position dependent opacity on a typical T Tauri disk is discussed in Sect.~\ref{sec:impact}. We also address two main factors that affect the ice features in our disk models, namely the relation between ice thickness and grain size, and the chemistry. We discuss the results of our models in the context of ice observations in Sect.~\ref{sec:discussion}. The method and results are summarized and concluded in Sect.~\ref{sec:conclusion}.

\section{Disk model and optical data}
\label{sec:prodimo}
{P{\small RO}D{\small I}M{\small O}\footnote{Version v2.0.0,  \url{https://prodimo.iwf.oeaw.ac.at/}}} or \underline{Pro}toplanetary \underline{Di}sk \underline{Mo}del \citep{woitke2009radiation,kamp2010radiation,thi2011radiation,rab2018x} is used in this paper to implement the position-dependent dust and ice opacities. The existing model, as explained in \citet{woitke2009radiation} and \citet{woitke2016consistent}, uses a fixed grain size distribution with no ices to calculate the opacities. The model starts by setting up the physical state of the disk such as gas and dust densities including dust settling of the bare grains. The code then calculates the dust opacities and solves the continuum radiative transfer, thereby determining the dust temperature structure.  Subsequently, the code computes the gas phase and ice chemistry and solves the heating \& cooling balance to find the gas temperature structure. We use steady state chemistry in all the models presented in this paper. One model also includes grain surface reactions. More details about the chemistry and the disk parameters are given in Appendix~\ref{app:typicalTT}.

The bare grain opacities are explained in \citet{woitke2016consistent}, based on \citet{min2016multiwavelength}. The opacity calculations assume that the grains are spherical and the materials are well mixed. The refractory dust grains are assumed to be composed of 60\% amorphous silicate (\citet{dorschner1995steps}, {$\rm Mg_{0.7}Fe_{0.3}SiO_3 $}) and 15\% amorphous carbon \citep{zubko1996optical}, with a porosity of 25\%.

The computation method presented in this paper requires multiple iterations to converge. The first iteration uses bare grain opacities. The resulting position-dependent ice abundance and position-dependent settled bare grain size distribution are used to recompute the dust opacities including the ices in the next iteration. Further, the code continues to iterate between radiative transfer, chemistry with ice formation, and opacity calculation, until convergence is achieved. The models presented in this paper converged within 3 such iterations.

The opacity calculations are based on Mie theory as explained in \citet{woitke2016consistent}. The dust grains (including those coated with ice) are considered to be spherical. Further, \citet{woitke2016consistent} use Mie theory and distributed hollow spheres \citep{2005A&A...432..909M} for calculating opacities. However, the grain size distribution at each grid point in the disk changes and the distributed hollow spheres are computationally expensive, in this paper distributed hollow spheres are not used. In this paper, the dependence of ice thickness on grain size is generalized and made computationally easy to apply to disk models. The porosity of the refractory grains and ice layer is fixed to a certain constant value, that is, the possible enhancement or diminution of porosity upon ice accretion, grain collision, shattering, and other thermal events is neglected.

\section{Opacity: Ice composition, thickness, and refractory materials}
\label{sec:opacity}
Extinction opacity ($\kappa_{\lambda}^{\rm ext}$) can be defined as the total absorption ($\kappa_{\lambda}^{\rm abs}$) and scattering ($\kappa_{\lambda}^{\rm sca}$) cross sections of the dust distribution, as shown in Eq.~(\ref{eq:opacity}) \citep{krugel2002physics}:

\begin{equation}
\begin{aligned}
    \kappa_{\lambda}^{\rm ext} &= \kappa_{\lambda}^{\rm abs} + \kappa_{\lambda}^{\rm sca}\\
    Q_{\lambda}^{\rm ext} &= Q_{\lambda}^{\rm abs} + Q_{\lambda}^{\rm sca}\\
    \kappa_{\lambda}^{\rm ext}&=\int_{a_{\rm min}}^{a_{\rm max}}\pi a^2 Q^{\rm ext}(a,\lambda) f(a)da
    \label{eq:opacity}\\
\end{aligned}
\end{equation}
where, $Q^{\rm ext}(a,\lambda)$, $Q^{\rm abs}(a,\lambda)$, $Q^{\rm sca}(a,\lambda)$ are the extinction, absorption, and scattering efficiencies defined as the ratio of respective cross sections to the geometric cross section of the grain, $a$ is the grain size, $a_{\rm min}$ and $a_{\rm max}$ are the minimum and maximum grain sizes, $f(a)$ is the considered size distribution ($=\frac{dN}{da}$, i.e., number of grains, $dN$, per size bin $da$).

Due to ice accumulating on the bare (refractory) grains, both extinction efficiency $Q(a,\lambda)$ and grain size distribution $f(a)$ change. These directly affect the opacity $\kappa_{\lambda}$. The effect of ice on these two factors are discussed in the following subsections.

\subsection{Effect of ice on size distribution}
\label{subSec:iceSizeDist}
The abundance of ice varies throughout the disk (see appendix). This means that the thickness of ice layer differs at each point in the disk. The thickness of ice on the grains at any point depends on the local volume of ice available and the total grain surface area available for ice accretion at that point in the disk. A simple relation can be written as shown in Eq.~(\ref{eq:iceLayerVol}).

\begin{equation}
    \begin{aligned}
        V_{\rm IG} &= V_{\rm I}+V_{\rm BG}\\
        n_{\rm d}\int_{a_{\rm min}}^{a_{\rm max}}\frac{4}{3}\pi(a+\Delta a)^3f(a)da&=V_{\rm I}+n_{\rm d}\int_{a_{\rm min}}^{a_{\rm max}}\frac{4}{3}\pi a^3f(a)da
    \end{aligned}
    \label{eq:iceLayerVol}
\end{equation}
where, $V$ is volume of material per unit volume of space with subscripts $\rm IG$ for icy grains, $\rm I$ for ice, and $\rm BG$ for bare grains, $a$ is the bare grain size, $\Delta a$ is the ice layer thickness, $f(a)$ [$\rm cm^{-1}$] is the normalized size distribution function of the grains, $a_{\rm min}$ and $a_{\rm max}$ are the minimum and maximum sizes of the bare grains, and $n_{\rm d}$ [$\rm cm^{-3}$] is the number density of grains. We consider the following two cases. (1) If, in Eq.~(\ref{eq:iceLayerVol}), $\Delta a$ is constant for all grain sizes, then the ice/refractory volume ratio ($V_{\rm I}$/$V_{\rm BG}$) is significantly larger for small grains than for large grains. (2) Ice thickness can also depend on the grain size. This can be due to coagulation or shattering of icy grains in the midplane or heating and re-condensation events \citep[the latter leads to growth of ice faster on larger grains,][]{kuroiwa2011evolution}. In this paper we assume that the limiting case of such scenario is an uniform ice/refractory volume ratio for grains of different sizes. 

\begin{figure}
    \centering
    \includegraphics[width=\linewidth, trim=15 5 140 95,clip]{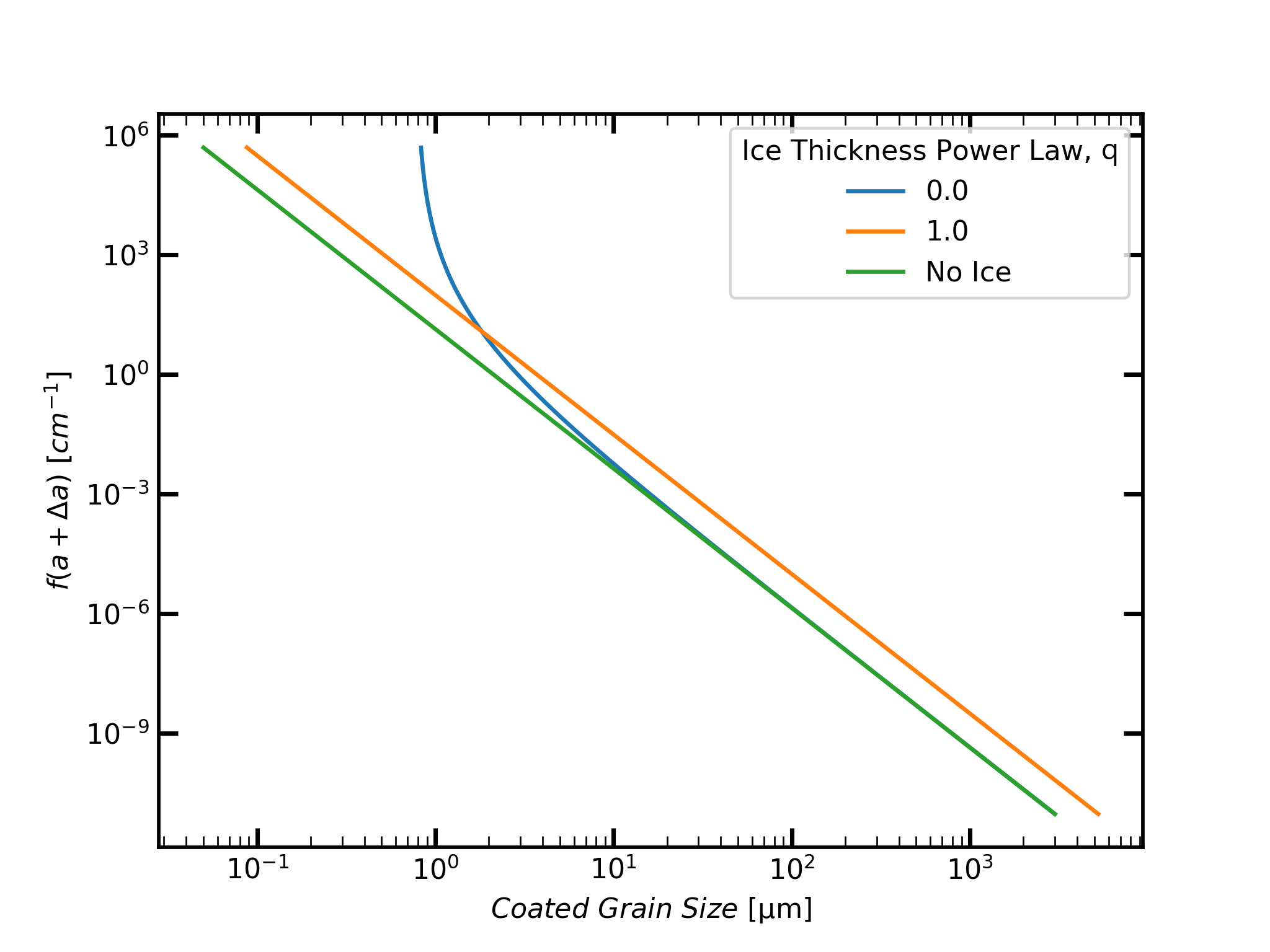}
    \vspace{-5mm}
    \caption{Icy grain size distribution for limiting values of ice coating powerlaw, $q$. Bare size distribution: $a_{\rm min}$ = 0.05 $\mu$m, $a_{\rm max}$ = 3mm, powerlaw index 3.5, $V_{\rm BG} = 6.2894\times 10^{-13}\rm\,cm^3/cm^3$, $V_{\rm I}=2.7577\times 10^{-12}\rm\,cm^3/cm^3$.}
    \label{fig:sizeDist}
\end{figure}

In order to allow for a general approach, which includes both limiting cases, we assume that the resulting grains are still spherical. For an icy grain of certain refractory size $a$, if $V_{\rm I}$, $V_{\rm BG}$, and $V_{\rm IG}$ are volumes of ice, refractory material, and total volume of the icy grain, respectively, then the ratio ${V_{\rm I}}/{V_{\rm IG}}$ and ${V_{\rm BG}}/{V_{\rm IG}}$ are the volume fractions of ice and refractory material in an icy grain respectively. If a grain with refractory size $a$ is present in an environment where there is no coagulation, shattering, or heating and recondensation events, then the ice thickness $\Delta a$ follows some constant $t$ and hence the volume fractions change across the size distribution. However, if the grain is present in an environment where the grain size distribution has been reached by coagulation, shattering, or heating and recondensation events, then the limiting case (according to the assumption) leads to an ice thickness $\Delta a$ proportional to the size of refractory material, $t\cdot a$, thus preserving the ice/refractory volume fraction. The following powerlaw approach is used to include both limiting cases
\begin{equation}
        \Delta a=t\cdot a^q
        \label{eq:icePowerLaw}
\end{equation}
where, $t$ and $q$ are arbitrary coefficients that are indicative of local ice abundance and the local dominance of shattering, coagulation, heating and recondensation events respectively. The two limiting cases are (1) $q\!=\!0$, where the thickness of the ice layer is constant and (2) $q\!=\!1$, where the ice/refractory volume fraction is constant and hence the thickness linearly depends on the bare grain size. The values $0<q<1$ relate to cases where the grain size distribution is affected by coagulation, shattering, or heating and recondensation events in varying levels. We assume that $f(a)$ in Eq.~(\ref{eq:iceLayerVol}) is the size distribution of the underlying bare grains. 

The effect of the ice covering the dust on the size distribution (dust + ice) is shown in Fig.~\ref{fig:sizeDist}. For a constant ice thickness ($q\!=\!0$), the relative increase in size of the smaller grains is several orders of magnitude and the new minimum grain size is determined by the ice layer thickness. The number density of grains of size close to the ice thickness increases significantly. However, the change in size for larger grain sizes is negligible. The overall effect on the grain size distribution is that it does no longer follows a powerlaw. For the other limiting case $q\!=\!1$, the small grains have small ice thicknesses and large grains have large ice thicknesses. Both minimum and maximum grain sizes are changed, but this is not as significant as in the case of constant ice thickness. Further, the total grain surface area increases, still dominated by smaller grains, but the size distribution powerlaw is retained.

The total volume of ice at any point in the disk $V_I$ is obtained from the chemical output of the disk code. If $n_{\rm i}\rm\,[cm^{-3}]$ is the number density at a grid point, $\mu_{\rm i}\rm\,[g]$ is the molecular mass, $\rho_{i}\rm\,[g/cm^3]$ is the material mass density of ice species $\rm i$, and $V_{\rm i}$ is the volume of ice species $\rm i$, then the total volume of ice per unit volume of space, $V_{\rm I}$, at a grid point is given by Eq.~(\ref{eq:iceVol}) for N ice species.
\begin{equation}
\begin{aligned}
    &V_{\rm i} = \frac{n_{\rm i}\mu_{\rm i}}{\rho_{\rm i}}\\
    &V_{\rm I} = \sum_{{\rm i}=1}^N  V_{\rm i}
    \label{eq:iceVol}
\end{aligned}
\end{equation}
Using Eq.~(\ref{eq:iceVol}), Eq.~(\ref{eq:iceLayerVol}) is solved numerically for $t$ at each grid point in the disk using Newton-Raphson method by approximating the root of the equation $V_{\rm IG}-V_{\rm I}-V_{\rm BG}=0$, for a given ice power law, $q$.

\subsection{Effect of ice on extinction, absorption, and scattering efficiency}
\begin{figure}
    \centering
    \includegraphics[width=\linewidth, trim=15 5 40 35,clip]{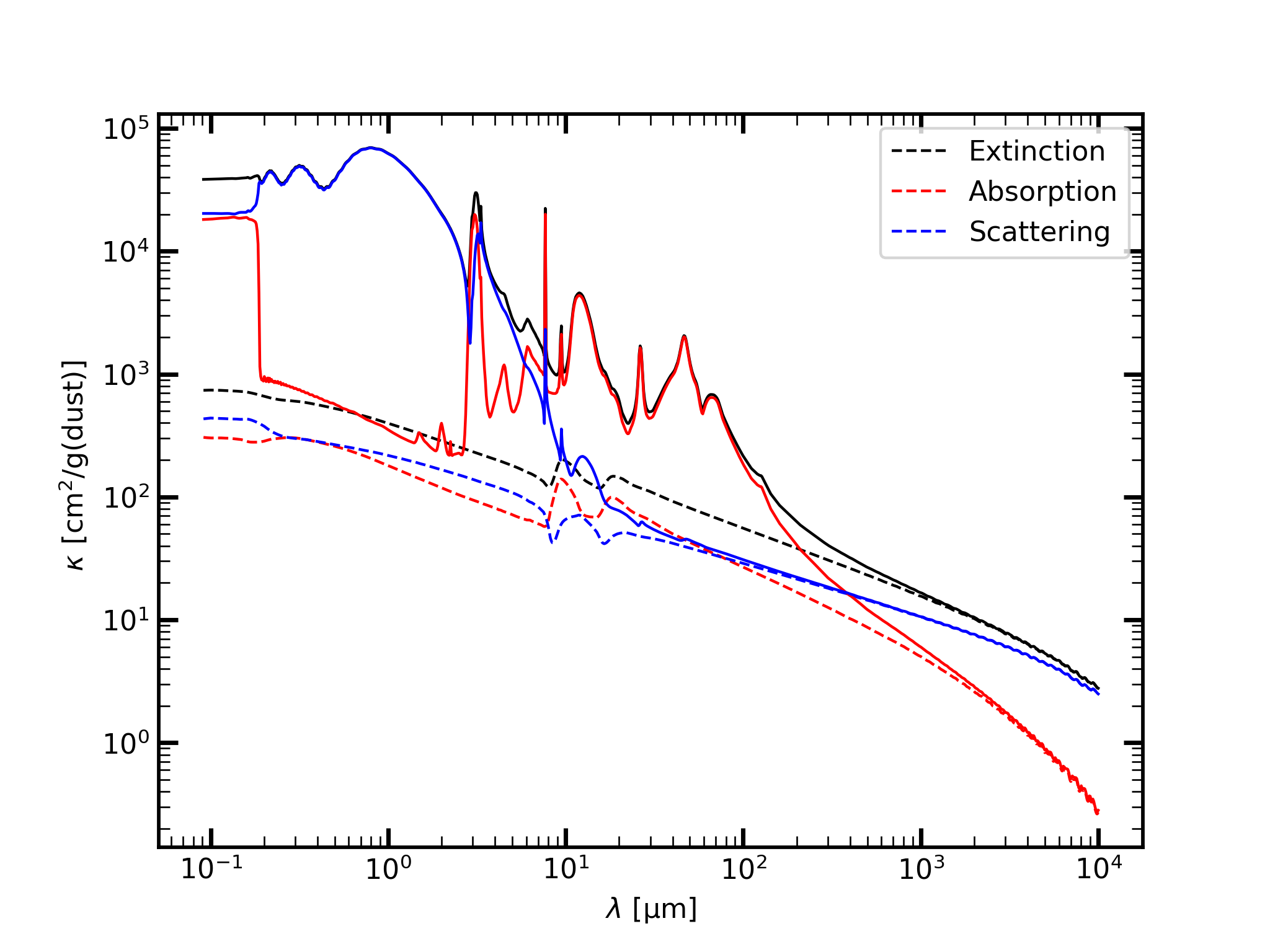}
    \vspace*{-4mm}
    \caption{Extinction, absorption, and scattering opacity of bare grains (dashed) and ice coated grains (solid). A constant thickness of ice layers is assumed for all grains ($q\!=\!0$). The composition of grains is $\rm Mg_{0.7}Fe_{0.3}SiO_3$, amorphous carbon, and porosity in the ratio 0.60:0.15:0.25, and that of ices: {$\rm H_2O$}, {$\rm CH_4$}, {$\rm NH_3$} in the ratio 0.60:0.20:0.20. The volume of dust is $V_{\rm BG} = 6.2894\times 10^{-13}\rm\,cm^3/cm^3$ and thickness of ice is $\Delta a = 0.246\,\mu m$. Refractory dust grain size distribution parameters are $a_{\rm min}\!=\!0.05\,\mu$m, $a_{\rm max}\!=\!3$\,mm, and powerlaw index 3.5. These opacities have been calculated using effective medium theory.}
    \label{fig:iceKappaEffect}
\end{figure}
Extinction ($Q_{\lambda}^{\rm ext}$), scattering ($Q_{\lambda}^{\rm sca}$), and absorption ($Q_{\lambda}^{\rm abs}$) efficiencies are related as given in Eq.~(\ref{eq:opacity}). These efficiencies are complex functions of size parameter ($x=2\pi a/\lambda$) and complex refractive index $m$($\lambda$).

The default Mie-efficiency calculations in {P{\small RO}D{\small I}M{\small O}}, as explained in Sect.~\ref{sec:prodimo}, assume that the same grain material is present everywhere in the disk (i.e.\ position-independent). In this work, however, at all locations in the disk where the local physical state allows for the presence of ice on grains, the opacity calculations include the increase of size and change of material composition of the grains due to ice formation. Two factors that should be considered for efficiency calculations are the different ice species and their abundances. The former affects the complex refractive index and the latter affects both the size and the refractive index of the icy grains. Figure \ref{fig:iceKappaEffect} shows the effect of these two factors on extinction, absorption, and scattering opacities by assuming a grain size distribution and fixed ice thickness ($q\!=\!0$) and composition. The figure shows the opacity cross sections per gram of refractory material [cm$^2$/g]. Scattering opacity increases both at shorter and longer wavelengths and absorption increases in the mid- to far-infrared range. This leads to an extreme albedo of >99\% at the optical to near-infrared wavelength region. Estimating the effects of these two factors, the different ice species and their abundances, on the Mie efficiencies at each grid point are individually explained in the following subsections.

\subsubsection{Effect of ice thickness on Mie efficiencies}
\begin{figure}
    \centering
    \includegraphics[width=\linewidth, trim=45 45 110 105,clip]{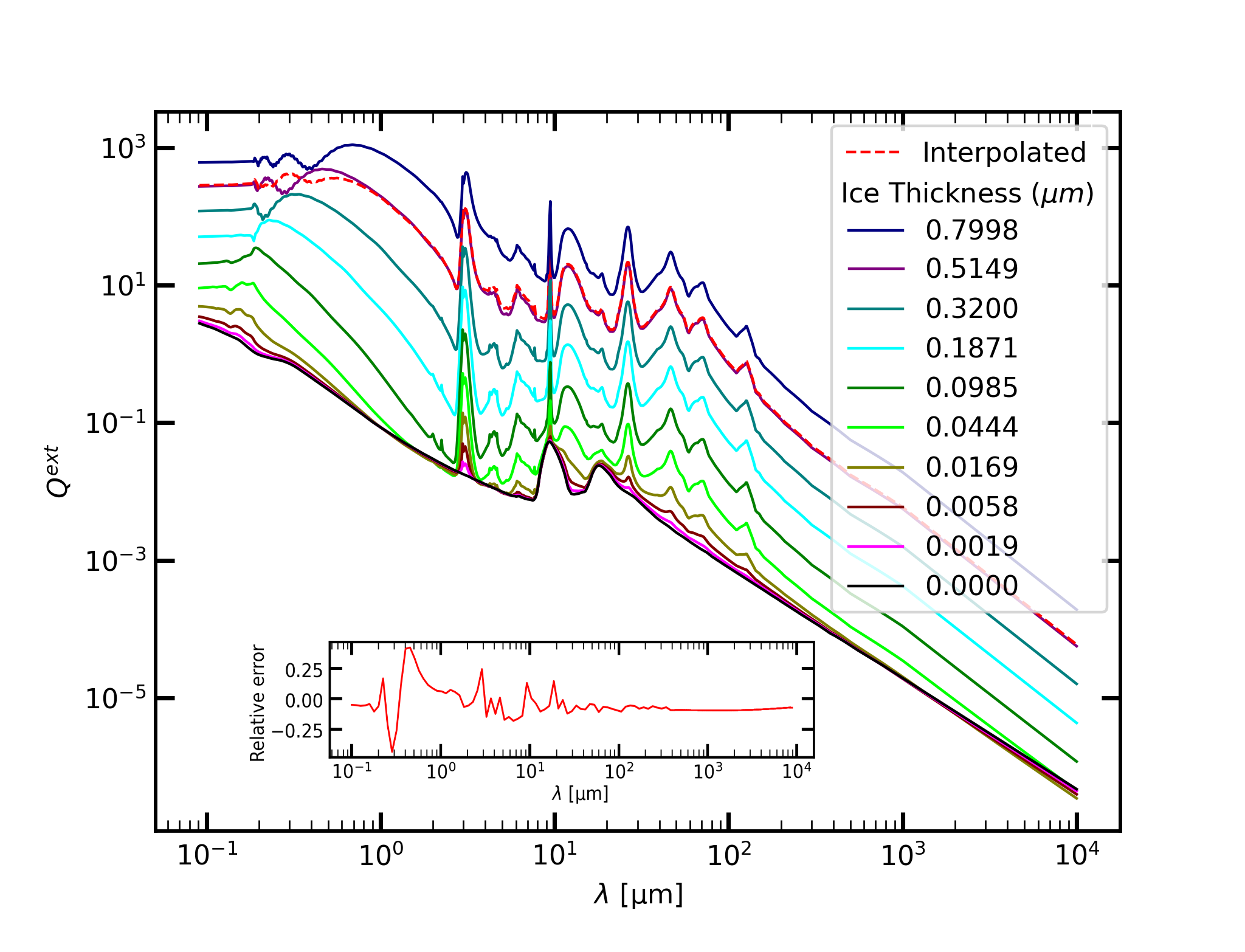}
    \vspace{-5mm}
    \caption{Extinction efficiency for a grain size 0.0559$\mu$m (with respect to the bare grain geometrical cross section) with different ice thicknesses composed of {$\rm H_2O$}, {$\rm NH_3$}, {$\rm CO$}, {$\rm CO_2$}, {$\rm CH_4$}, {$\rm CH_3OH$} in ratio 0.45:0.20:0.10:0.05:0.05:0.15 with 20\% of porosity. Also shown is  the validity of the interpolation for a constant ice thickness layer of 0.515\,$\mu$m. The embedded panel shows the relative error between the interpolation and explicit calculation of efficiency for the icy grain.}
    \label{fig:iceThickInterpol}
\end{figure}
To estimate the position dependent opacities, a Mie efficiency matrix ($Q$-matrix) is first generated. Each point in this multi-dimensional matrix corresponds toMie efficiency defined by individual ice species, ice thickness, ice power law, bare grain size, and wavelength. For a given bare grain size distribution and the wavlength bins, a finite number of ice thickness and ice power law values (Eq.~\ref{eq:icePowerLaw}) for each ice species are used to populate this matrix. Figure \ref{fig:iceThickInterpol} shows how extinction efficiency changes as the thickness of ice layer increases. Since it is computationally expensive to calculate the efficiencies based on thickness at each grid point using Mie routines, log interpolation (Eq.~\ref{eq:logInterpol}) between points in the pre-calculated $Q$-matrix is done.

In Eq.~(\ref{eq:logInterpol}), $\theta$ and $\phi$ are the interpolation weights which define the closeness of the local thickness to the nearest two thickness grid points. However, for interpolation between the grid points of the smallest ice thickness and bare grain opacities, a simple linear interpolation is used. Figure \ref{fig:iceThickInterpol} presents the values obtained by such interpolation. It should be noted that in Fig.~\ref{fig:iceThickInterpol}, the efficiency, $Q_{\rm ext}$, is the ratio of extinction cross section of ice covered grains to the geometrical cross section of bare grains. Here, bare grain geometrical cross section is considered because it makes the log interpolation of efficiency simple. 
\begin{equation}
    \begin{aligned}
        &Q_{\rm a,\lambda}(r,z)=Q_{\rm a,\lambda}(t_1)^{\theta}\cdot Q_{\rm a,\lambda}(t_2)^{\phi}\\
        &\theta=\frac{log(t_{\rm 2}/t_{\rm r,z})}{log(t_{\rm 1}/t_{\rm 2})}\ \ \ \phi=\frac{log(t_{\rm 1}/t_{\rm r,z})}{log(t_{\rm 1}/t_{\rm 2})}
    \end{aligned}
    \label{eq:logInterpol}
\end{equation}
where, $\rm r,z$ are the radial distance from the rotation axis and vertical distance from the midplane of a point in disk, $t_{\rm r,z}$ is the thickness coefficient (see Eq.~\ref{eq:icePowerLaw}) of ice layer at the location $\rm (r,z)$ in the disk for which the opacity is being estimated, $t_{\rm 1}$ and $t_{\rm 2}$ are the two ice thickness values considered in the pre-calculated $Q$-matrix closest to $t_{\rm r,z}$, using which the efficiencies $Q_{\rm a,\lambda}(t_{\rm 1})$ and $Q_{\rm a,\lambda}(t_{\rm 2})$ are obtained from the matrix, and $\theta$, $\phi$ are the weights for log interpolation. Finally, the values calculated from Eq.~(\ref{eq:logInterpol}) are scaled by $(a$/$\Bar{a})^2$ to obtain the actual efficiencies with respect to ice covered grains ($\Bar{a}$ is size of the grain with ice. i.e. $\Bar{a}=a+\Delta a=a+t_{\rm r,z}\cdot a^{q_{\rm r,z}}$). However, the local ice thickness power law ($q_{\rm r,z}$) requires extensive solutions of shattering, coagulation physics, and heating-recondensation events, which cannot be obtained by disk chemistry. Hence, trivial values ($q$=0,1) are considered in Section \ref{sec:impact} independent of position (i.e. same throughout the disk).

\subsubsection{Effect of ice composition on Mie efficiencies}
The protoplanetary disks are generally dominated by {$\rm H_2O$} ice \citep{oberg2011spitzer}. Though the direct observation of ices in disks are scarce, observations of molecular clouds and young stellar objects (YSOs) show the presence of {$\rm H_2O$}, {$\rm CO$}, {$\rm CO_2$}, {$\rm NH_3$}, {$\rm CH_3OH$}, {$\rm CH_4$} amongst other ice species \citep{gibb2000inventory,gibb2004interstellar,boogert2008c2d,pontoppidan2008c2d,noble2013survey}. These six ice species are considered for opacity calculations in this paper. Our sources for the refractive indices are listed in Table~\ref{tab:iceList}. In this paper, the composition of ice at any point in the disk (the number density of ice species, see Eq.~\ref{eq:iceLayerVol}) is obtained from the chemistry results of the disk code, which is explained in \cite{kamp2017consistent}. The ice composition is determined by accretion and desorption processes (including thermal, UV, and cosmic ray processes). The effect of grain surface chemistry is discussed in Section \ref{sec:impact}.

\begin{figure}
    \centering
    \includegraphics[width=\linewidth, trim=15 5 40 35,clip]{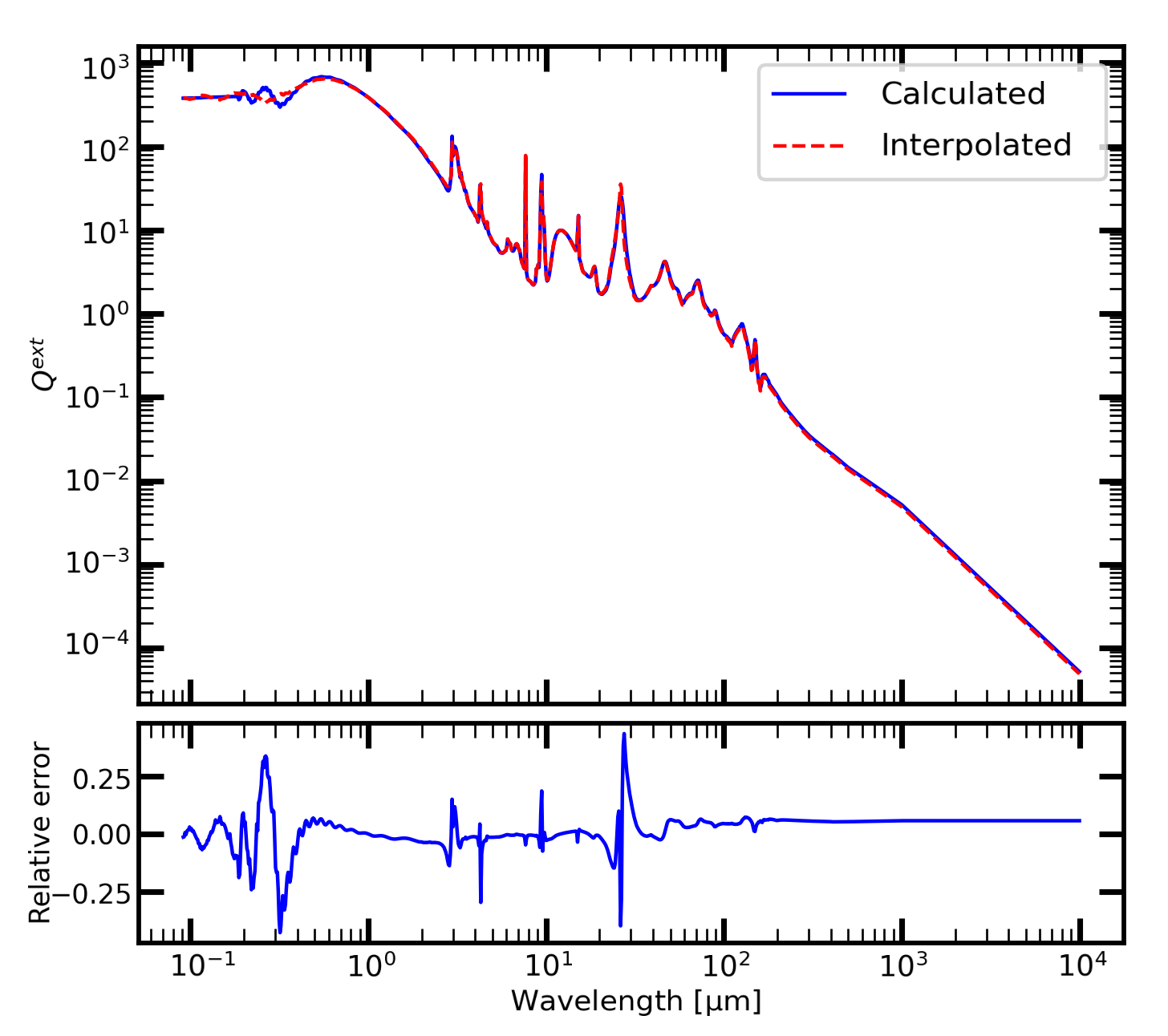}
    \vspace*{-5mm}
    \caption{Extinction efficiency for a grain size 0.0559$\mu$m with 0.515$\mu$m ice thickness composed of {$\rm H_2O$}, {$\rm NH_3$}, {$\rm CO$}, {$\rm CO_2$}, {$\rm CH_4$}, {$\rm CH_3OH$} in ratio 0.16:0.16:0.16:0.2:0.16:0.16 with 20\% of porosity. The bottom panel shows the relative error between the interpolation and explicit calculation of efficiency for the icy grain. }
    \label{fig:iceSpeciesInterpolation}
\end{figure}

It is found that the efficiencies based on local ice composition, for a given ice thickness and a given bare grain size distribution, can be approximated by a simple volume fraction weighted summation of efficiencies from the $Q$-matrix for grains coated with different pure ices. This is shown in Eq.~(\ref{eq:iceSpeciesInterpolation}). This method is computationally efficient because the number of Mie calculations required becomes independent of the spatial grid size of the disk, instead depends on the number of ice species, $N$, considered.
\begin{equation}
    Q_{\rm a,\lambda}({\rm Bare,Ice_1,Ice_2,...,Ice}_n)=\sum_{i=1}^N\left(\eta_i Q_{\rm a,\lambda}({\rm Bare+Ice}_i)\right)
    \label{eq:iceSpeciesInterpolation}
\end{equation}
where, $Q_{\rm a,\lambda}$ is the efficiency, the left hand side of the equation is the efficiency of the grains coated with local mixed ice, composed of species $1$ to $N$, $\eta_i$ is their respective ice abundance volume fraction ($\eta_{\rm i}=V_{\rm i}/V_{\rm I}$, Eq.~\ref{eq:iceVol}). It should be noted that in Eq.~(\ref{eq:iceSpeciesInterpolation}), the $Q_{a,\lambda}$ on left and right side of the equation are based on the same refractory grain composition and size. Further, $Q_{a,\lambda}$ on the right is calculated using the same thickness of ice as in the left except that the ice layer is assumed to be entirely composed of $\rm Ice_i$. The validity of this approach is shown in Fig.~\ref{fig:iceSpeciesInterpolation}, which compares the efficiencies calculated for grain coated with a mixed ice layer (1) by using effective medium theory for ice mixture and (2) by weighted summation as given in Eq.~(\ref{eq:iceSpeciesInterpolation}). The relative error plotted in the bottom panel of the figure shows that the interpolation is in good agreement with calculation using effective medium Mie routine. These approximations (Eqs.~\ref{eq:logInterpol} and \ref{eq:iceSpeciesInterpolation}) introduces uncertainties (of similar order as the relative errors) in the ice features in the disk spectra obtained from the solution. However, these uncertainties are $\sim$1 order of magnitude smaller than the uncertainties introduced due to difference in the opacities of a mixed and a differentiated core-mantle grain. Hence, these approximations are sufficiently accurate for our models but the uncertainties have to be taken into account while fitting observations. The following section discusses the effective medium and core-mantle Mie theory.

\subsection{Effective medium vs core-mantle Mie theory}
The composition of the icy grains can be thought of as (1) a mixture of ices and refractory material or (2) a multi-layered structure with refractory material as the core and an ice mantle on top. The opacity calculations for these two cases can be carried out using either Mie theory with effective medium theory or core-mantle Mie theory, respectively. However, the computational effort required for core-mantle solutions is a factor 4 larger than the other.

\begin{figure}
    \centering
    \includegraphics[width=\linewidth, trim=40 40 140 115,clip]{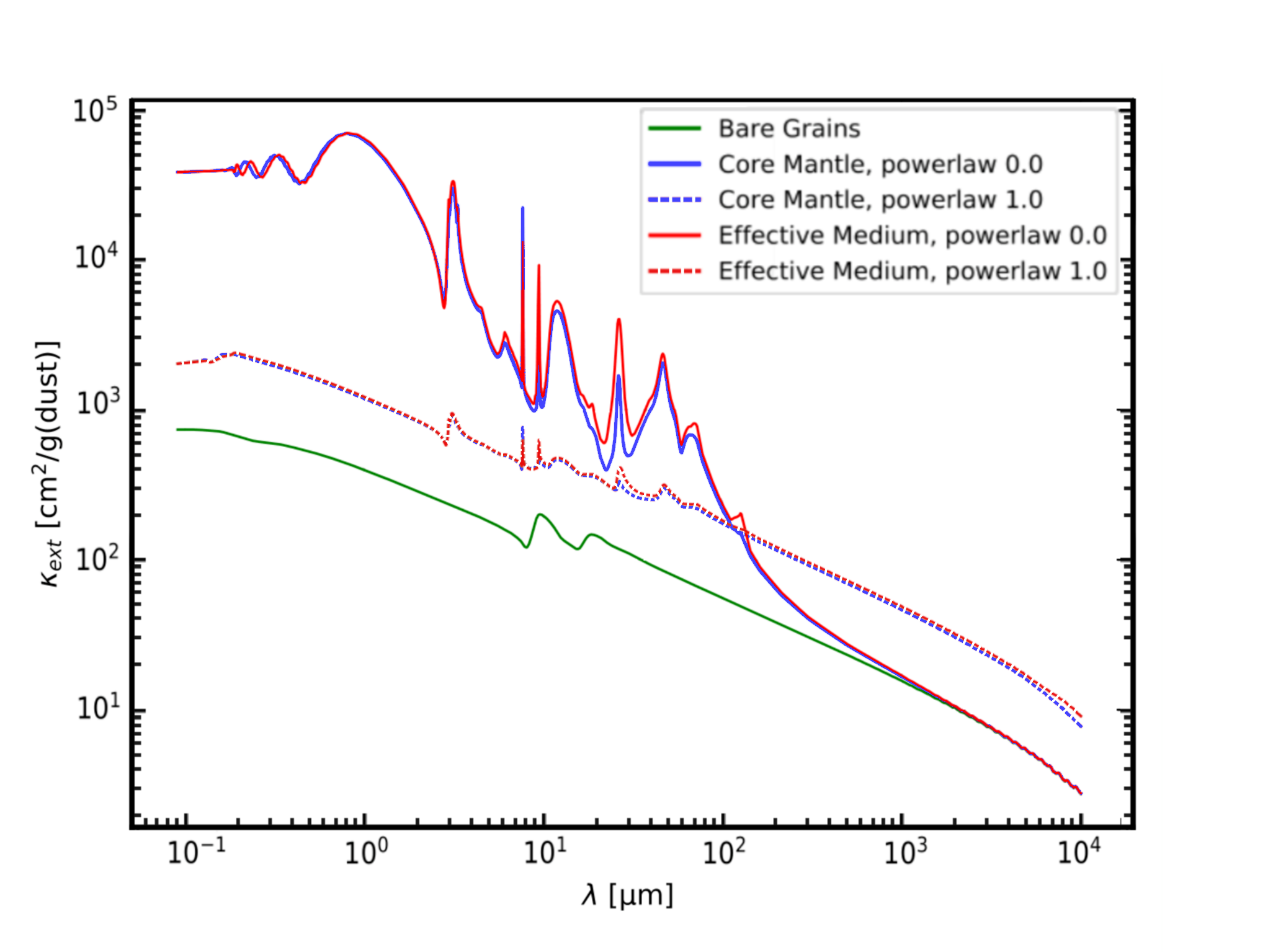}
    \vspace*{-5mm}
    \caption{Extinction opacity of bare grains (green) and ice covered dust grains calculated using the effective medium theory (red) and Core-Mantle Mie theory (blue) for ice powerlaws $q\!=\!0$ (solid lines) and $q\!=\!1$ (small dashed lines). The ice composition is {$\rm H_2O$}:{$\rm NH_3$}:{$\rm CH_4$}:Vacuum = 0.51:0.40:0.04:0.05. The bare grain size distribution, composition, and volume of ice is the same as in Fig.~\ref{fig:iceKappaEffect}}
    \label{fig:EM_CM}
\end{figure}

\begin{figure*}
    \centering
    \includegraphics[width = 0.9\linewidth]{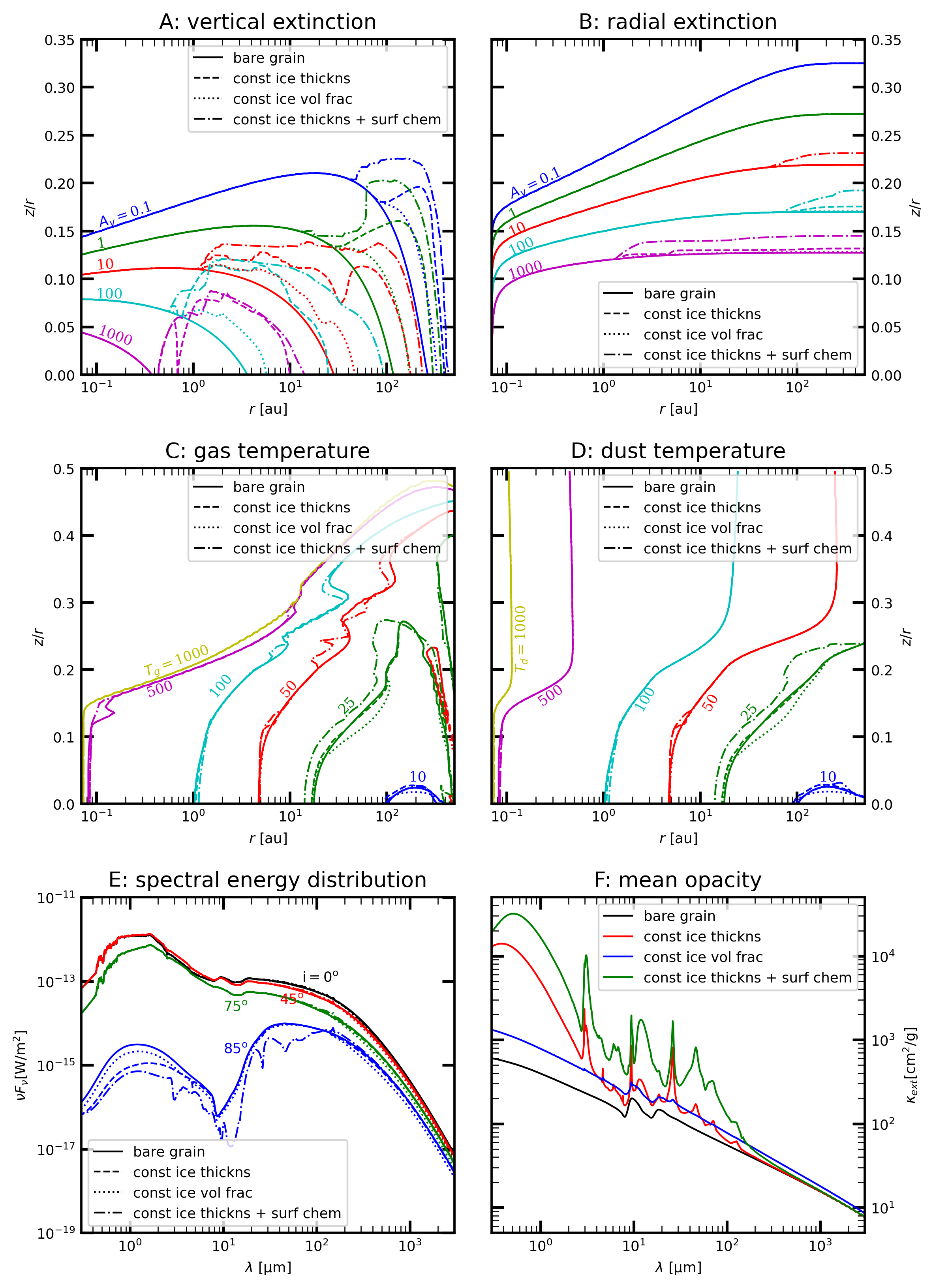}
    \caption{Comparison of disk properties of the models. Panels \textbf{A-F} show the vertical visual extinction, radial visual extinction, dust temperature, gas temperature, spectral energy distribution, and disk mass averaged opacity of the entire T\,Tauri disk [in $\rm cm^2/g$(refractory material)] respectively for four disk models. The four models are 1) bare grains, 2) icy grains with constant thickness of ice layer ($q\!=\!0$), 3) icy grains with constant ice/refractory volume ratio ($q\!=\!1$), 4) icy grains with constant ice thickness ($q\!=\!0$) with grain surface reactions. All the models use steady state chemistry.}
    \label{fig:Combined}
\end{figure*}

\begin{figure*}
    \centering
    \includegraphics[width = 0.96\linewidth]{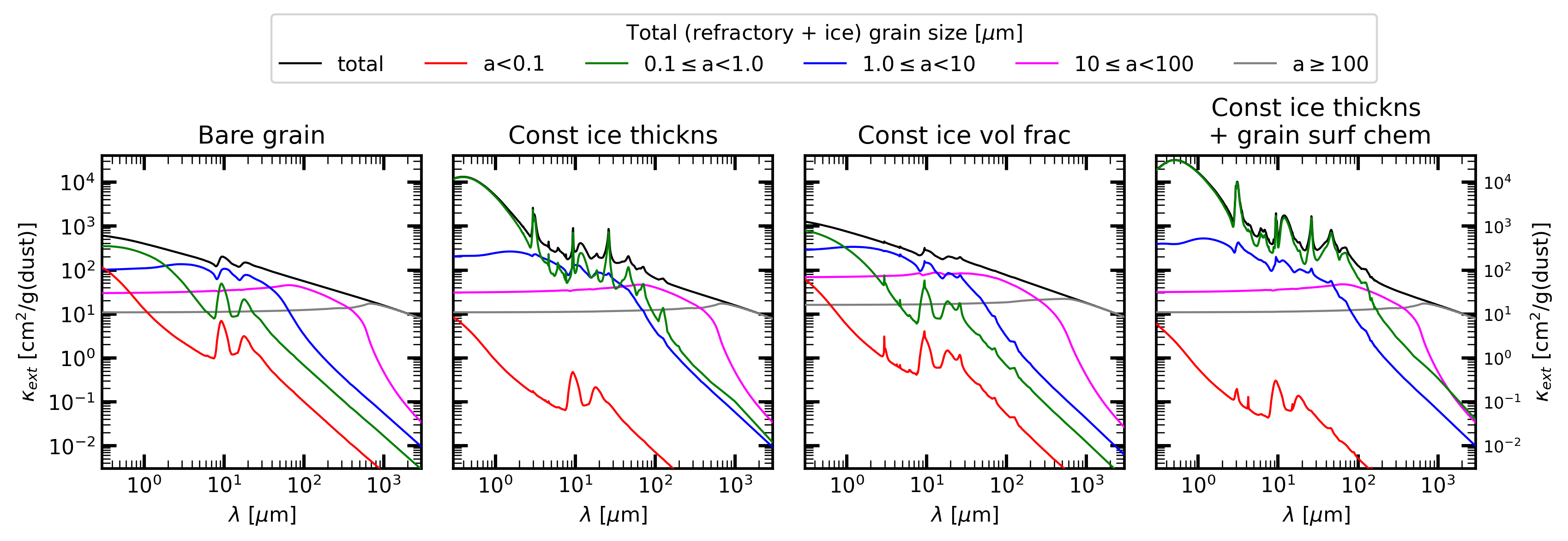}
    \caption{Contribution of different grain sizes (refractory + ice) to the disk mean opacity for bare grain model, constant ice thickness model, constant ice volume fraction, and constant ice thickness with grain surface chemistry model. The legend at the top indicates the curves corresponding to different grain size bins.}
    \label{fig:opacComponents}
\end{figure*}

It can be argued that in the limiting case $q\!=\!1$, the coagulated or shattered grains do not retain a core-mantle structure. Further, it is debatable whether icy grains are layered or mixed in dense environments \citep{2017MNRAS.467.4753N}. For example, ices could be organized as layers starting with a layer of polar ices such as {$\rm H_2O$}, {$\rm NH_3$}, and {$\rm CH_3OH$} and on top a layer of non-polar ices such as {$\rm CO$}, N$_2$, and O$_2$, creating an onion-like structure \citep{oberg2009complex}. Ice covered grains can undergo processing due to thermal processes, ultraviolet photons, and cosmic ray particles, called energetically processed ice \citep{boogert2015observations}, which can mix the ices in the mantle and as well cause segregation. Impinging energetic particles and irradiation with X-rays and UV photons may cause ice desorption \citep{palumbo2006cosmic,andersson2008photodesorption}, mantle explosion \citep{ivlev2015impulsive}, or the formation of new ice species \citep{1988Icar...76..225A,caro2002amino,palumbo2006cosmic}, possibly followed by molecular ejections \citep{Dulieu2013} which can lead to mixed ices. The mixed ice components can undergo processing leading to ice restructuring and segregation \citep{fayolle2011laboratory,oberg2009quantification}. Further, thermal and radiative processes can also form larger and more complex molecules \citep{garrod2008complex,gerakines1996ultraviolet,oberg2009formation,caro2002amino}. In the present model, neither ice segregation nor restructuring is considered. Moreover, using Eq.~(\ref{eq:iceSpeciesInterpolation}) does not take into account the ice features that result from interactions between different ice species, for example {$\rm CO$}-{$\rm H_2O$} at 4.68\,$\mu$m, 4.647\,$\mu$m \citep{tielens1991interstellar,sandford1988laboratory}, {$\rm NH_3$}-{$\rm H_2O$} at 3.4~$\mu$m \citep{1980JChPh..73.4832B}. Figure \ref{fig:EM_CM} shows the ice opacities calculated for the limiting ice powerlaws with effective medium theory and core-mantle Mie theory (here, the mantle is a composed of a mixture of ices). The differences between the opacity values calculated by these two approaches are small. Due to these reasons our opacity calculations are based on effective medium theory.

\section{Impact of ice opacities on the disk models}
\label{sec:impact}
In the following, we discuss four protoplanetary disk models with identical stellar, disk shape and mass, dust settling, and bare grain parameters, see Appendix~\ref{app:typicalTT}. The models only differ with respect to the treatment of the ice opacities and grain surface chemistry. All four models use steady state chemistry and the fourth model includes grain surface chemistry. The four models are:
\begin{itemize}
    \item bare grain: bare grain opacity. 
    \item const ice thickns: ice opacities with a constant ice thickness, $q\!=\!0$ (Eq.~\ref{eq:icePowerLaw}).
    \item const ice vol frac: ice opacities with a constant ice volume fraction, $q\!=\!1$ (grain size dependent ice thickness). 
    \item const ice thicknss + surf chem: same as the constant ice thickness model, that is, $q\!=\!0$, but includes grain surface chemistry. 
\end{itemize}
The position-dependent icy grain opacity calculations are based on the abundances of ice species as provided by the chemical network at each point in the disk. The iteration between chemistry, opacity calculation, and continuum radiative transfer, as described in Sect.~\ref{sec:prodimo}, is continued until the opacity computations converge to a tolerance of less than 5\%.

\subsection{Extinction and opacity}
The inclusion of the ice opacities significantly changes the extinction properties of the disk in the model, particulary in the midplane, where ices are abundant. Fig.~\ref{fig:Combined} shows the vertical visual extinction, radial visual extinction, dust temperature, gas temperature, spectral energy distribution, and disk mass averaged opacity of the entire T\,Tauri disk [in $\rm cm^2/g$(refractory material)] for the four disk models. The assumptions of constant ice thickness ($q\!=\!0$) and constant ice volume fraction ($q\!=\!1$) modifies the grain size distribution differently (Fig.~\ref{fig:sizeDist}), which in turn affects the mean disk opacity. Panel F of Fig.~\ref{fig:Combined} shows that including ice opacities significantly increases the disk mean opacity compared to the bare grain opacity (black) model. The constant ice thickness model (red) shows an increase in the opacity at shorter wavelengths (<3~$\mu$m) by more than an order of magnitude. At longer wavelengths (>100~$\mu$m), the opacity change with respect to bare grain model is negligible. The mean opacity of the constant ice volume fraction model (blue) is larger than the bare grain model at all wavelengths. Particularly, at shorter wavelengths the increase in opacity is by a factor $\sim$2 and at longer wavelengths by a factor of $\sim$1.3. At intermediate wavelengths, the opacities of both constant ice thickness and constant ice volume fraction models are of the same order of magnitude. However, the constant ice volume fraction model produces weaker ice features. The model with constant ice thickness and grain surface chemistry (green) shows that an increase in the ice abundance (due to grain surface reactions) leads to a larger opacity compared to all the other models in the mid-IR wavelengths. For a given volume of ice at one location in a disk, the effect of the ice thickness to grain size relation on the opacity can also be seen in Fig.~\ref{fig:EM_CM}. The difference in the opacities between the models is an interplay of changing grain size distribution due to the ice thickness to grain size relation and the absolute abundance of ice due to chemistry which is explained in the following paragraph.

The model with constant ice thickness produces ice features that are stronger than for the model with constant ice volume fraction by an order of magnitude. However, the ice abundances are similar in these models. This implies that ice distribution across the grain sizes become important for the ice features. Figure~\ref{fig:opacComponents} shows the contribution of different grain sizes to the disk mean opacity for the four models. Grains of particular sizes ($a$) have opacity maxima at wavelengths regions corresponding to $2\pi a/\lambda\sim1$. At wavelengths corresponding to $2\pi a/\lambda>>1$, the Mie extinction efficiencies reach a constant value ($Q_{ext}\to 2$). At wavelengths corresponding to $2\pi a\lambda<<1$, the extinction is dominated by absorption ($Q_{ext}\simeq Q_{abs}$) and falls off inversely proportional to the wavelength ($Q_{abs}\propto \lambda^{-1}$) \citep{krugel2002physics}. This behavior is seen in the bare grain model (left). For the assumed refractory grain size distribution, the opacity in the wavelength range 2~$\mu$m-100~$\mu$m is dominated by grains of sizes 1.0~$\mu$m$\leq$a$<$10~$\mu$m and 10~$\mu$m$\leq$a$<$100~$\mu$m. Further, the small grains (<1~$\mu$m) have strong silicate (and ice for other models) features. In the constant ice thickness model, small grains (<1~$\mu$m) have large volume fractions of ice ($\sim$90\%) due to larger surface area. This leads to change in grain size distribution as shown in Fig.~\ref{fig:sizeDist}, where there is a pile up of grains around the size corresponding to ice thickness and removal of grains smaller than the ice thickness. Due to these changes, the opacity contribution of grains smaller than local ice thickness drastically reduces ($\sim$1 order) and that of grains of size corresponding to the ice thickness drastically increases ($\sim$1 order) (Fig.~\ref{fig:opacComponents}). That is, the main opacity carriers in the infrared wavelengths changes from 1.0~$\mu$m$\leq$a$<$100~$\mu$m to 0.1~$\mu$m$\leq$a$<$10~$\mu$m. This effect is more pronounced in the constant ice thickness model with grain surface chemistry (right) in which the relevant grain sizes are 0.1~$\mu$m$\leq$a$<$1~$\mu$m. In contrary, in the constant ice volume fraction model most of the ice resides on larger grains (>100~$\mu$m) which do not contribute to opacity in the near- and mid-infrared wavelengths and the change in size distribution is uniform across grain sizes. Thus, models with constant ice thickness show stronger ice features. The opacity curve of the bare grain model shows two silicate features at 10 $\mu$m and 20 $\mu$m. The constant ice volume fraction model shows small features around 2.95~$\mu$m ({$\rm NH_3$}), 3~$\mu$m ({$\rm H_2O$}), 4~$\mu$m ({$\rm CO_2$}), 26~$\mu$m ({$\rm NH_3$}), and the silicate features. Due to the changes in the size distribution, the models with constant ice thickness (with and without grain surface chemistry) show significantly enhanced ice features, prominently showing {$\rm H_2O$}, {$\rm NH_3$}, {$\rm CO_2$}, {$\rm CO$}, and {$\rm CH_4$}. Hence, the average opacity of the disk strongly depends on 1) the ice power law, $q$, in Eq.~(\ref{eq:icePowerLaw}) and 2) total abundance of ices.

Panels A and B show the effect of the ice opacities on the radial and vertical visual extinctions. When ice opacities are included, the vertical extinction increases significantly toward the midplane, but only once the disk has become optically thick. This increase is more than an order of magnitude in the models that use a constant ice thickness (in which the opacity increases significantly as well) closer to the midplane, like an opacity wall or barrier. For example, the vertical visual extinction at 30\,au in the midplane changes from about 10 to 100 when $q\!=\!0$ ice opacities are considered. The observable disk extent can be defined as the vertical optical depth be equal to unity at optical wavelengths. This observable disk extent increases from about 120\,au for the bare grain model to about 300\,au for the models with constant ice thickness. This behavior is mostly a size effect, because ice formation increases the sizes of the small grains in particular, which are the carriers of the opacity at optical wavelengths. However, the impact of the icy opacities in the $q\!=\!1$ model is less pronounced. Hence, the visual extinction profile in a disk strongly depends on the relation between the ice thickness and grain size, that is, the ice power law index $q$ in Eq.~(\ref{eq:icePowerLaw}).

\begin{figure*}
    \centering
    \includegraphics[width=\linewidth]{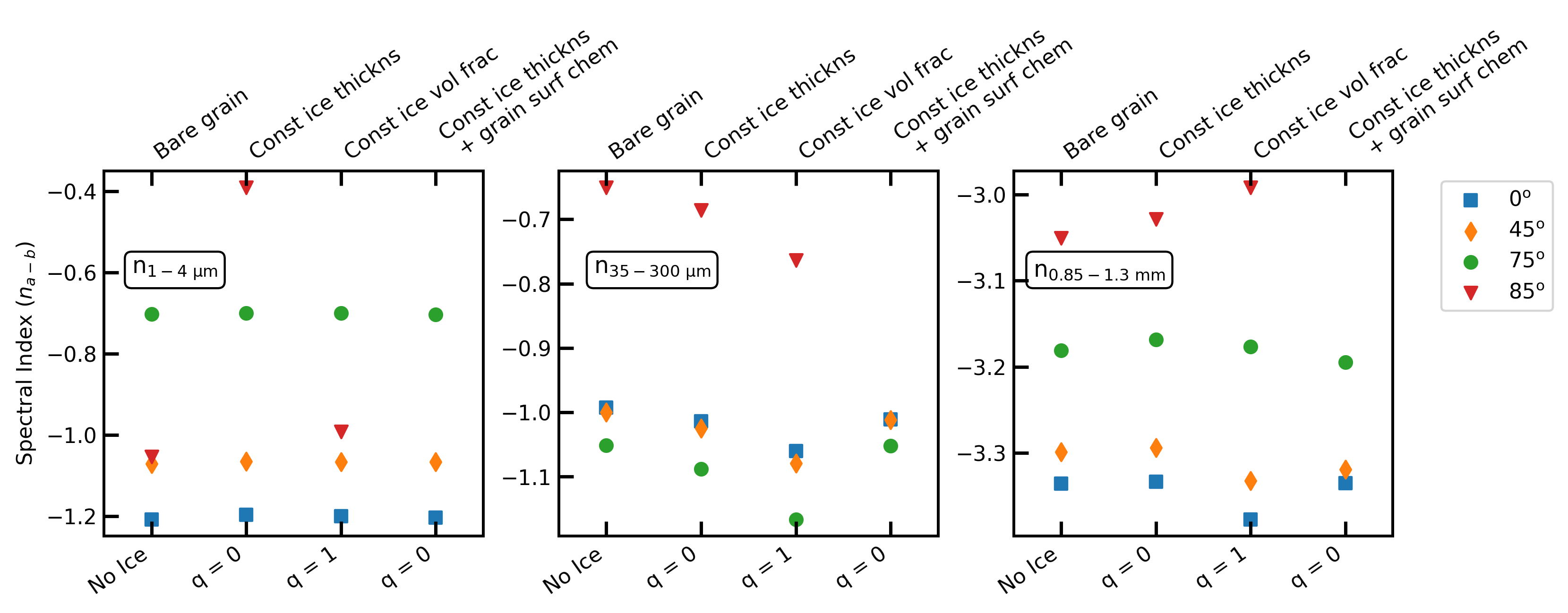}
    \caption{Spectral indices for $\rm 1-4\ \mu m$ (left), $\rm 35-300\ \mu m$ (middle) and $\rm 0.85-1.30\ mm$ (right) for the four models presented in Sect.~\ref{sec:impact}. Different markers represent the spectral indices for different disk inclinations as indicated in the legend on the right.}
    \label{fig:SEDslope}
\end{figure*}

\subsection{Spectral energy distribution and temperature profile}
\label{sec:modelSEDTemp}
Although the mean opacities in the disk increase drastically when ice coating is included, the main disk properties such as temperature, pressure, and chemical abundances, etc., do not change significantly in the models.  This is because the ice only builds up in the (radially and vertically) optically thick and shielded regions in the disk. In particular, there is almost no ice in the model at locations where stellar or interstellar UV photons can penetrate, even if the local dust temperatures are low. This is because of the UV photodesorption process, which removes ice species as they interact with UV photons. The changes of opacity in the optically thick regions have no significant influence on the local mean intensities ($J_{\nu}$), as $J_\nu \approx B_\nu(T_{\rm d})$ is valid anyway ($B_{\nu}$ is the Planck or black-body radiation), and hence there is no significant influence on the radiation that escapes from these regions. Panels C and D of Fig.~\ref{fig:Combined} show the computed dust and gas temperatures in the four disk models. The surface layer of the disk re-emits or scatters radiation received from the star toward the observer and toward the disk midplane \citep{natta1993temperature}. The latter largely determines the temperatures in the disk \citep{chiang1997spectral,chiang2001spectral}. Since the ice is formed in regions that are already optically thick, there is almost no ice in the upper regions of the disk that can scatter incoming radiation. Thus, the temperature structure does not change much when position dependent ice opacities are included. For example, the largest change in temperature at any grid point compared to the bare grain model is ~18\%; this is for the model with grain surface chemistry. The overall temperature changes are less than 1.5\% for the models without grain surface chemistry and less than 5\% for the model with grain surface chemistry.

\begin{figure*}
    \centering
    \includegraphics[width=0.8\linewidth]{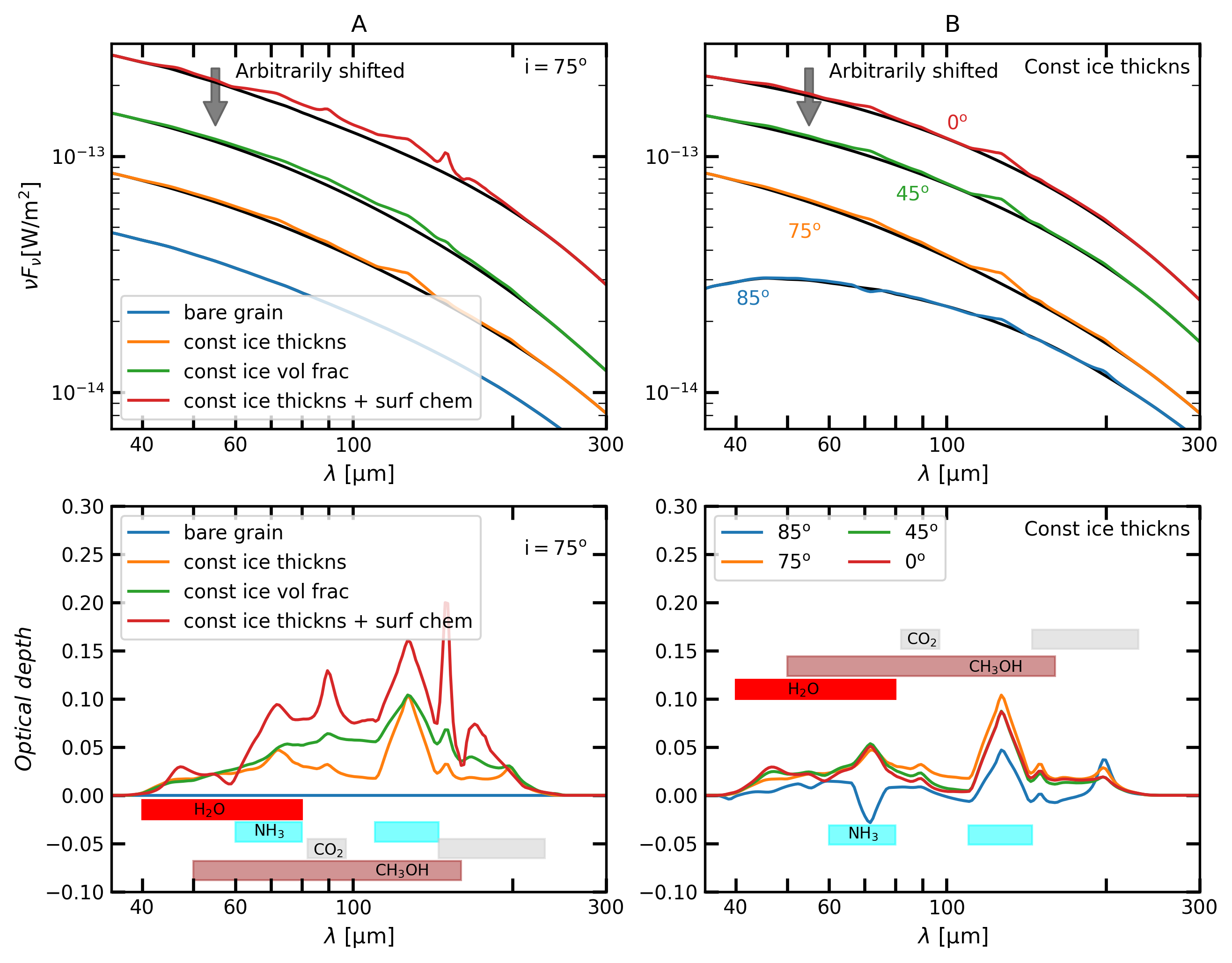}
    \caption{Spectral energy distributions and optical depths of ices in different models. The top panels show the spectral energy distributions of different models between $\rm 40\ \mu m$ and $\rm 300\ \mu m$. The black curves indicate a pseudo-continuum used to obtain the optical depths of the ice features. In each panel, the SED of the different models are arbitrarily shifted to provide better visualization. The bottom panels show optical depths of the ice features in the SEDs in the top panels. Panel A shows the ice features in the four models at a disk inclination of 75${\rm ^o}$. Panel B shows the ice features for the model with constant ice thickness without grain surface chemistry at different disk inclinations. Note that the orange curve in each panel is the same model. The ice features are marked using colored rectangles in the optical depth plots.}
    \label{fig:emissionFeatures}
\end{figure*}

The spectral energy distributions (SEDs) of the four models are presented in panel E of Fig.~\ref{fig:Combined} for disk inclinations 0$\rm ^o$, 45$\rm ^o$, 75$\rm ^o$, and 85$\rm ^o$. Since the visual extinction and the albedo in the disk surface, and the temperature in the disk largely remain unaffected, we do not see substantial change in the shape of the SED. However, at higher inclinations due to large optical depths in the midplane, the flux density at shorter wavelengths (<3$\mu$m) for the models with constant ice thickness is about an order of magnitude smaller compared to the bare grain model. Similarly, the increase in opacity at longer wavelengths for the model with constant ice volume fraction (for example, the opacity at 10$^{\rm3}\,\mu$m is 1.17 times that of bare grain model) results in slightly lower flux density in the longer wavelength region of the SED. This is encapsulated in Fig.~\ref{fig:SEDslope} which shows the spectral indices ($\rm n_{a-b}$) for $\rm 1-4\ \mu m$, $\rm 35-300\ \mu m$, and $\rm 0.85-1.3\ mm$ for the four models at different disk inclinations. The spectral index is defined by the following relation:
\begin{equation}
    n_{a-b} = \frac{ log_{10}(\lambda_a F_a)-log_{10}(\lambda_b F_b)}{log_{10}(\lambda_a)-log_{10}(\lambda_b)}
\end{equation}
where $\lambda F$ is the flux density and $\lambda$ is the wavelength. The three spectral indices probe different regions of the disk. n$_{\rm 1-4~\mu m}$ corresponds to the hot inner disk surface, n$_{\rm 0.85-1.3~mm}$ corresponds to the cooler outer disk midplane, and n$_{\rm 35-300~\mu m}$ corresponds to warm disk regions. The spectral range for the latter is large because multiple ice features are present between these wavelengths and determination of the slope becomes difficult for smaller wavelength ranges, see Fig.~\ref{fig:emissionFeatures}.

The spectral indices of the models at near- to mid-infrared wavelengths (n$_{\rm 1-4~\mu m}$) remain constant across the models for a given inclination. This is because the radiation at these wavelengths is emitted from the surface layers close to the inner rim of the disk. These regions are devoid of ices due to thermal desorption. At millimeter wavelengths, the radiation arises from large part of the disk and hence can be used to estimate the dust mass, size of the grains, and their evolution in the disk \citep{2007prpl.conf..767N,2014prpl.conf..339T}. Thus, spectral indices at millimeter wavelengths (n$_{\rm 0.85-1.3~mm}$) are used to quantify the dust properties \citep[e.g.][]{2019PASP..131f4301W,2020A&A...642A.164V}. The spectral indices for any given model depends on the inclination (Fig.~\ref{fig:SEDslope}). In other words, the model with the same dust properties exhibit different mm-spectral indices at different inclinations. Secondly, for a given disk inclination, the spectral index at millimeter wavelengths can depend on the ice opacity and chemistry model. Since the radiation at these wavelengths arise from the bulk of the outer and cooler disk, the disk mass averaged opacity can be used to understand the behavior of the spectral indices. Panel~F of Fig.~\ref{fig:Combined} shows that the model with a constant ice volume fraction has a steeper opacity curve at these wavelengths and the model with a constant ice thickness has essentially the same opacity curve as the bare grain model. This is also reflected in the spectral indices at lower inclinations (0{$\rm^o$}, 45{$\rm^o$}). However, at higher inclinations, the orientation effects play a role leading to certain parts of the disk being less visible and some being more visible. At mid- to far-infrared wavelengths, the spectral index (n$_{\rm 35-300\mu m }$) becomes more negative (steeper slope) for models with ice opacities without grain surface chemistry. The model with constant ice volume fraction has a lower spectral index than the model with constant ice thickness at all disk inclinations. This shows that the ice power law has an effect on the slope of mid- to far-infrared SED. The model with constant ice thickness with grain surface chemistry shows the spectral indices similar to that of the model without grain surface chemistry. The emission region corresponding to these wavelengths spans across large range of temperatures (due to large wavelength range), hence it is difficult to qualitatively analyze the complex interaction of ice opacities and the spectral index. The indices for the model with grain surface chemistry for disk inclination of 85$^{\rm o}$ are unreliable due to substantial ice features, hence are not shown. The ice power law determines the change in grain size distribution (Fig.~\ref{fig:sizeDist}) and the distribution of ice across the grain sizes. Thus, the ice power law influences the slope of the SED at longer wavelengths.

\begin{figure*}
    \centering
    \includegraphics[width=0.95\linewidth]{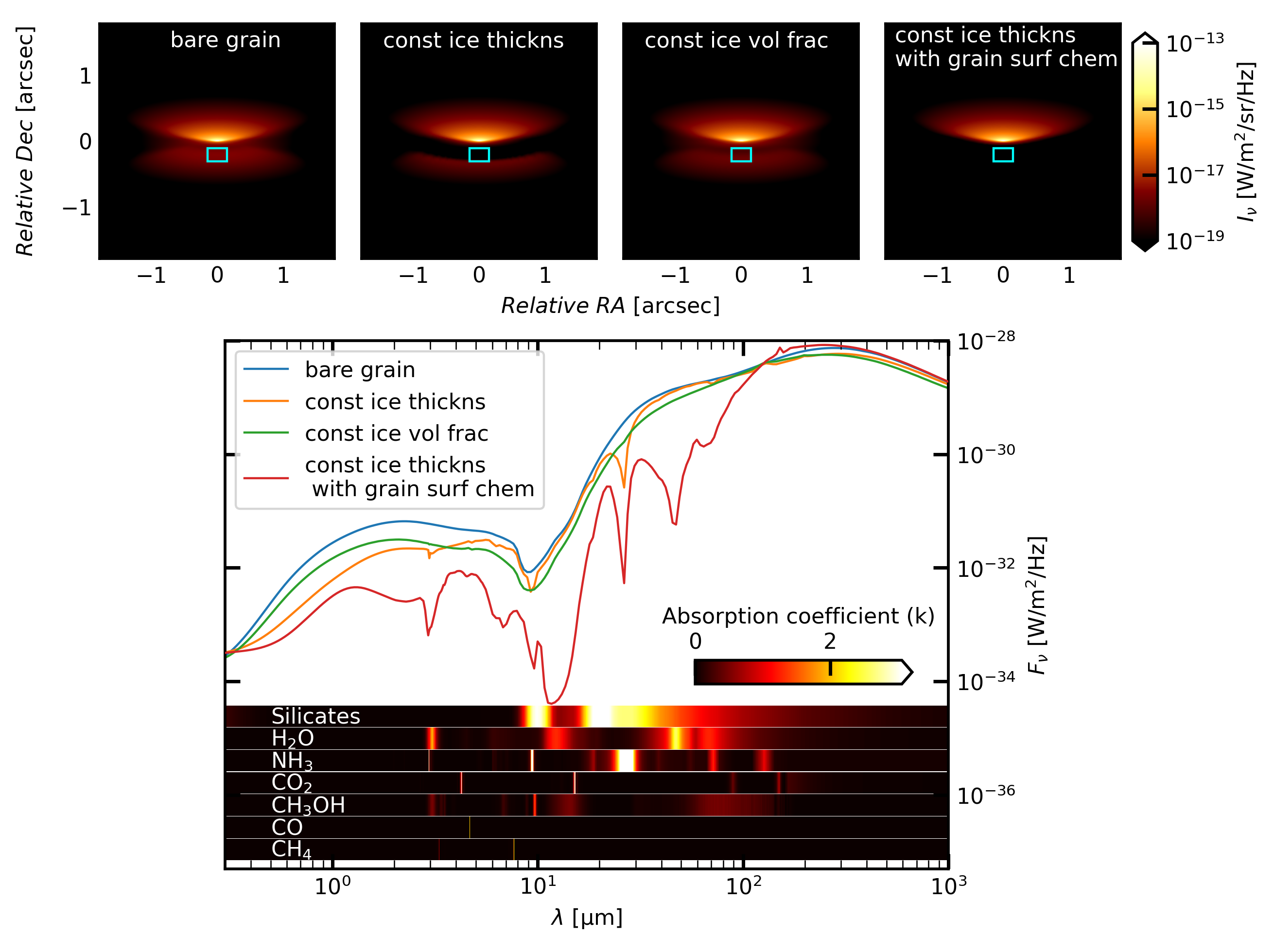}
    \vspace*{-4mm}
    \caption{Simulated images and spatially resolved spectral flux densities of the disk models presented in this work. The top panels show simulated images (intensity maps) at 26\,$\mu$m and inclination of 75$\rm ^o$. The cyan rectangles (0.2$''\times$ 0.3$''$) indicates the region whose spectral flux density is plotted in bottom panel. The spectral flux density plot also shows the absorption coefficients of the ices and silicates.}
    \label{fig:CompareImages}
\end{figure*}

\subsection{Ice emission and absorption features}
\label{sec:observation}

\begin{figure}
    \centering
    \includegraphics[width = \linewidth]{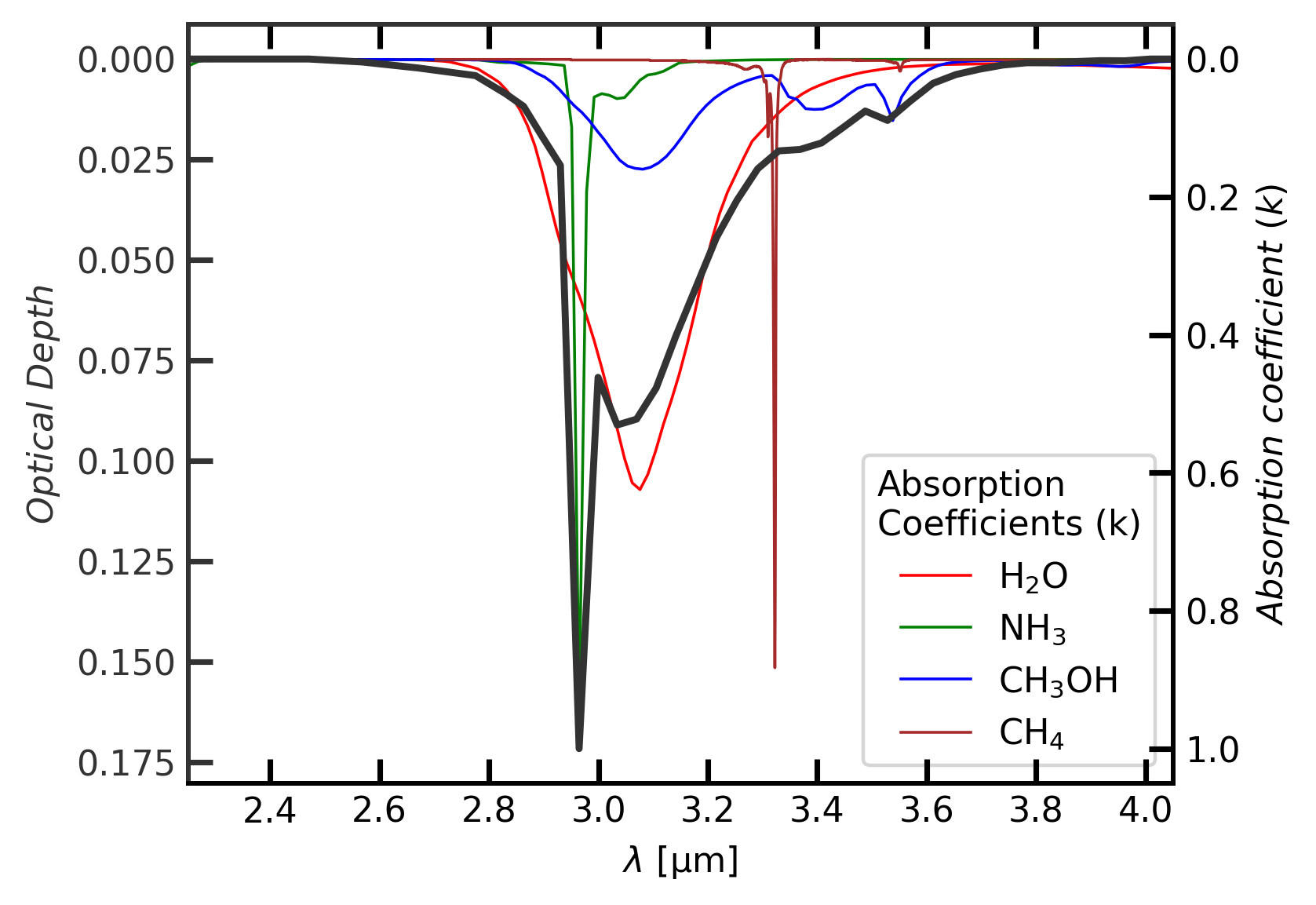}
    \caption{Optical depth of the 3~$\mu$m {$\rm H_2O$} ice feature derived from the spectral flux density obtained from model 2 in Fig.~\ref{fig:CompareImages}. The optical depth is shown in dark gray thick line and the absorption coefficients of ice species that have a feature in the plotted wavelength range are shown as colored thin lines.}
    \label{fig:OpticalDepth}
\end{figure}

The models produce ice absorption and emission features in the SED. At inclinations lower than the edge-on case, the SEDs do not show any ice absorption features in any of the four models. However, weak emission features at far-infrared wavelengths can be observed corresponding to {$\rm H_2O$}, {$\rm NH_3$} ice, and {$\rm CO_2$} ice. Close to edge-on ($\rm 85^o$), models with constant ice thickness show ice absorption features. Particularly, the model without grain surface chemistry shows only a strong {$\rm CO$} ice absorption feature and a weak {$\rm NH_3$} ice absorption feature at 4.6~$\mu$m and 26~$\mu$m respectively. The model with grain surface chemistry shows significantly more ice absorption features (and stronger) in addition to those seen in previous model, such as {$\rm H_2O$}, {$\rm CO_2$}, and {$\rm CH_3OH$}. This model also shows stronger ice emission features at far-IR wavelengths. This shows that the chemical network strongly affects the strength of the ice absorption and emission features. Therefore, observations of such ice features can lay strong constraints on the type of chemistry in disks.

Figure~\ref{fig:emissionFeatures} shows the zoomed-in SED (top panels) and optical depths (bottom panels) of the ice emission features (at $\rm 75^o$ inclination) from $\rm 40-300\ \mu m$ for the models discussed before (Sect.~\ref{sec:impact}). The black curves in the SED plots in all the top panels show the pseudo continuum used to obtain the optical depths of the ice features. This continuum is obtained by spline interpolation using python library \textit{scipy} \citep{2020SciPy-NMeth} and then correcting spline of other models for systematic error produced by the spline for bare grain model. The top and bottom plots in Panel~A show the SED and optical depth of the four models. Due to higher ice abundances, the model with grain surface chemistry produces stronger features (by about a factor 2 in optical depth) than the model without grain surface chemistry. The peaks of the optical depths of emission features for the constant ice thickness and the constant ice volume fraction models are quite similar, however, the constant ice volume fraction model shows larger optical depths (by a factor less than 2) between the peaks of these features. The figure also shows the ices that correspond to the features, the prominent ices being {$\rm H_2O$}, {$\rm NH_3$}, {$\rm CO_2$}, and {$\rm CH_3OH$}. Unlike other ices, {$\rm CH_3OH$} ice has a broad feature that spans from about 50~$\mu$m to 160~$\mu$m and hence does not show up as a single peak instead contributes to the optical depth across the wavelength range. Panel B of Fig.~\ref{fig:emissionFeatures} shows that the optical depths are quite similar across the inclinations except for the close to edge-on case, where some of the features (such as $\rm 47~\mu m$, $\rm 70~\mu m$ {$\rm H_2O$} and {$\rm NH_3$} ice features) can be seen in absorption. Thus, the strength of the ice features depend on the chemistry producing the ices, the relation governing ice distribution over different grain sizes and the disk inclination.

The SED shows absorption features only at high disk inclinations. Hence, it is interesting to explore the observability of absorption features when the disk is spatially resolved. Figure~\ref{fig:CompareImages} shows images of the four models at 26~$\mu$m in the top panels. The bottom of the figure shows the spectral flux density of these models obtained from the region marked by the cyan rectangles. The cyan rectangles are $\rm 0.2''\times 0.3''$ in size\footnote{which can be similar to binning 6 pixels of JWST NIRSpec IFU pixels (which are $\rm 0.1''\times 0.1''$ each)}. This is a simple study to estimate the fluxes for spatially resolved observations and is not an attempt to simulate any specific instrument. Hence, point spread function, noise, errors, sensitivity, and saturation limits are not considered. Simulating observations and identifying best observation strategies are left to a future work.

The images in the top panels show that inclusion of ice opacities produces a distinct band of absorption around the midplane, subsequently called the dark lane. This dark lane is more prominent for models with constant ice thickness and also depends on the chemical network. These differences reflect the increase in the extinction in the disks as discussed in Sect.\,\ref{sec:impact}. Panel~A of Fig.~\ref{fig:Combined} shows that the inclusion of ice opacities increases the extinction radially, more strongly for models with constant ice thickness compared to model with constant ice volume fraction. The ice power law has a strong effect on the appearance of the disks, particularly on the size of the dark lane. The spectral flux density (in the bottom panel) extracted in the dark lane shows that ice absorption features can be seen by spatially resolving the disks at inclinations less than the edge-on case. Further, the constant ice volume fraction model shows substantially weaker features compared to the models that consider a constant ice thickness. The absorption coefficients of ices and the refractory silicate used are also shown at the bottom of the figure to help identify the features observed in the spectra. More information regarding the optical constants used in this paper is described in Appendix~\ref{app:iceOpConst}. The models show features corresponding to five of the six ices included in the opacity calculation. Though the local abundance of {$\rm CH_4$} ice can be as large as $1.4\cdot10^{-4}$ in the disk, no feature corresponding to it is observed. This is because, though the local {$\rm CH_4$} abundance can be large, the column density along the line of sight of the spectra extracted in Fig.~\ref{fig:CompareImages} for other ices such as {$\rm H_2O$}, {$\rm NH_3$} and {$\rm CH_3OH$} are larger due to the spatial extent of these ices (see appendix). Moreover, the {$\rm CH_4$} features lie in the wavelength regions with features from more dominating ices. This can also be seen in the disk mean opacity of the model with grain surface chemistry as shown in panel F of Fig.~\ref{fig:Combined}. The model with grain surface chemistry shows stronger absorption features than the model without grain surface chemistry by about 2 orders of magnitude. Ice emission features are also present at longer wavelengths but are substantially weaker (by several orders of magnitude) compared to the absorption features. Hence, spatially resolving the disk produces significant ice absorption and emission features that are not seen in the disk integrated spectra, particularly strong in the absorption by few orders of magnitude.

Figure~\ref{fig:OpticalDepth} shows the optical depth of the $\rm 3\ \mu m$ ice feature obtained from spatially resolved spectral density plot of the model with constant ice thickness from Fig.~\ref{fig:CompareImages}, along with the absorption coefficients of different ice species that have a feature in that wavelength range. The 3~$\mu$m profile is a combination of {$\rm H_2O$}, {$\rm NH_3$}, and {$\rm CH_3OH$} ices. The sharp peak at 2.95~$\mu$m in the profile is due to {$\rm NH_3$} ice. {$\rm H_2O$} ice dominates the profile of the 3~$\mu $m feature, however, the longer wavelength side (>3.3~$\mu$m) shows a contribution of {$\rm CH_3OH$} ice. NIRSpec IFU should be able to spatially resolve the disks and hence we can expect observations of such features from JWST.

\section{Discussion}
\label{sec:discussion}
Ice accretion on the grains modifies the grain size distribution. In Sect.~\ref{subSec:iceSizeDist} we showed that a constant ice thickness leads to a deviation of the size distribution from the powerlaw and a constant ice volume fraction increases the minimum and maximum grain sizes. Further, these changes can be enhanced by models that produce more ice, such as the model with grain surface chemistry. The emission at (sub) millimeter wavelengths probe the bulk of the dust in the disk. This corresponds to the large dust grains which are settled in the disk midplane. In these environments grain evolution processes such as grain growth are important as they affect the emission at these wavelengths. \citet{2007prpl.conf..767N} show that the millimeter slope is strongly affected by the maximum grain size and the dust distribution powerlaw. Particularly, when the maximum grain size is of the order of millimeters, the slope is very sensitive to the changes in maximum grain size. Since, the emission in these wavelengths can be used to quantify the dust properties and their evolution, it is important to consider the effect of ice on the grain size distribution and the opacity while interpreting these spectral indices.

Ices have been detected in protoplanetary disks, as discussed in Sect.~\ref{sec:intro}. These detections are from the total disk integrated fluxes (SED) and are not spatially resolved. Our models show weak emission features at far-IR wavelengths at all inclinations, which can be, however, very weak and barely observable for some model setups. Absorption features can only be seen in the edge-on orientation in the total disk spectra. Even when seen edge-on, most of the absorption features are seen only when grain surface reactions are included, for example, the $\rm 3\ \mu m$ {$\rm H_2O$} feature. However, spatially resolving the disk produces a multitude of absorption features across the spectral range.

In contrary, the $\rm 3\ \mu m$ water ice feature has been reported in total disk flux of some objects (mostly edge-on) such as HK Tau B \citep{terada2007detection}, d132-1832,  d216-0939 \citep{terada2012adaptive}. Further, emission features are observed in a few protoplanetary disks \citep[one such example is VW Cha,][]{mcclure2015detections}. The presence of emission features and absence of absorption features in our models can be attributed to the limited parameters explored. Objects differ, for example in terms of the parent star, disk mass, the inclination, the flaring angle, inner and outer disk radii, gaps, silhouette disks, etc. All these parameters can strongly affect the ice features. Consequently, each disk has to be modeled separately and our models are not representative of every disk. Further, dynamic physical processes such as vertical and radial transport are not included in our models and can significantly alter the ice features. For example, vertical mixing can significantly change the ice composition in the disk \citep{2014ApJ...790...97F} and readily produces ice features that are stronger compared those from our models by factors of a few in terms of optical depth (Woitke et al. in prep). Similarly, models with gaps can also show strong ice features in the SED. The goal of this paper is to present the consistent position dependent opacities, and exploring such parameters and simulating observations is left to future work. For example, the disks with ice features in the far-IR reported by \cite{mcclure2015detections} can be modeled with our ice opacity approach.

\citet{B21} also modeled ices in disks and they find absorption features of multiple ices in the total disk flux even at inclinations lower than the edge-on case. However, these models differ from our models significantly and hence a direct comparison is not possible. For example, \citet{B21} do not consider the interstellar UV field and use a cosmic ray ionization rate that is an order of magnitude lower than our models. These assumptions can significantly affect the ice features. The vertical extent of ice in the disk is largely determined by the interstellar UV field, because this radiation can photodesorb the ices back into the gas phase. This in turn inhibits the presence of ice in the emitting regions of the disk. The ice destruction mechanism by photodesorption is consistently treated in our models, with chemistry-dependent ice opacities and fully coupled UV radiative transfer. Further, \citet{B21} use two dust populations with significantly different settling conditions and the treatment of ice opacities is not fully consistent with local conditions in each grid point of the model.

The strengths of the ice features strongly depend on the ice power law and the chemistry network. Spatially resolving the disk presents multiple ice features and can probe the vertical and radial distribution of ice in the disk. In the end, future observations will allow us to test these various models and to refine the assumptions.

\section{Conclusions}
\label{sec:conclusion}
A computationally efficient way of computing position-dependent ice and bare grain opacities in protoplanetary disks has been implemented in {P{\small RO}D{\small I}M{\small O}}, which consistently couples chemistry, ice formation, opacity computations, and radiative transfer in protoplanetary disks. The conclusions drawn are the following:
\begin{enumerate}
    \item A weighted sum of the opacities of grains coated with the different pure ices and log interpolation of opacities between pre-calculated points in ice thickness space provides a computationally efficient and sufficiently accurate method to calculate the icy opacities in protoplanetary disks.
    \item Locally, the continuum opacity can change significantly by ice formation. The opacity changes are particularly large at shorter wavelengths, due to the increased size of the small grains, when assuming a constant ice thickness on all grains. Opacity changes are less pronounced, and more uniform in wavelength space, when the grains are assumed to have a constant ice/refractory volume ratio, which is expected if coagulation and shattering are important. In both cases, the primary reason for those opacity changes is the altered size distribution of the icy grains.
    \item The position-dependent ice opacities do not alter the physical disk structure significantly, because the ice opacities are significant only in the optically thick disk regions ($A_V\!\ga\!10$) in our models.
    \item The spectral energy distributions (SEDs) of the disks show faint ice emission features in the far-infrared regions beyond 35~$\mu$m for all our models with ice opacities. However, absorption features are seen only at high inclinations, close to edge-on conditions. In particular, the 3~$\mu$m water ice feature is seen only in case of the model with grain surface chemistry in close to edge-on orientation.
    \item The assumption made on how the ice is distributed across the grain size distribution (ice power law) influences the far-infrared and millimeter slope of the SED.
    \item The strengths of the ice features are influenced by the ice power law (about a factor of 2 in optical depths). The number of ice features and their strength increase (a factor 2 in optical depth) in models having grain surface chemistry, which is directly related to an increase in ice abundance. 
    \item Our models show a prominent dark lane in resolved images. This dark lane is caused by icy grains situated just above the disk midplane, and the size of the dark lane depends on the assumptions used to calculate the modified dust size distribution and the chemical network.
    \item To detect the ice absorption features, extracting spatially resolved spectra from the dark lane is more promising.
\end{enumerate}
The models presented here are useful to model existing (e.g. Herschel) and upcoming observations (e.g. JWST) to better understand the ice composition in disks. And further improvements, such as transport processes, will help us understand the role of these processes in disk environments.
\begin{acknowledgements}
CHR acknowledge the support of the the Deutsche Forschungsgemeinschaft (DFG, German Research Foundation) - 325594231. CHR is grateful for support from the Max Planck Society.
\end{acknowledgements}

%
  \bibliographystyle{aa} 
  \bibliography{aa.bib} 
%

\begin{appendix}
\section{Ice optical constants}
\label{app:iceOpConst}
\begin{figure}[!ht]
    \centering
    \includegraphics[width=\linewidth, trim=20 5 15 35,clip]{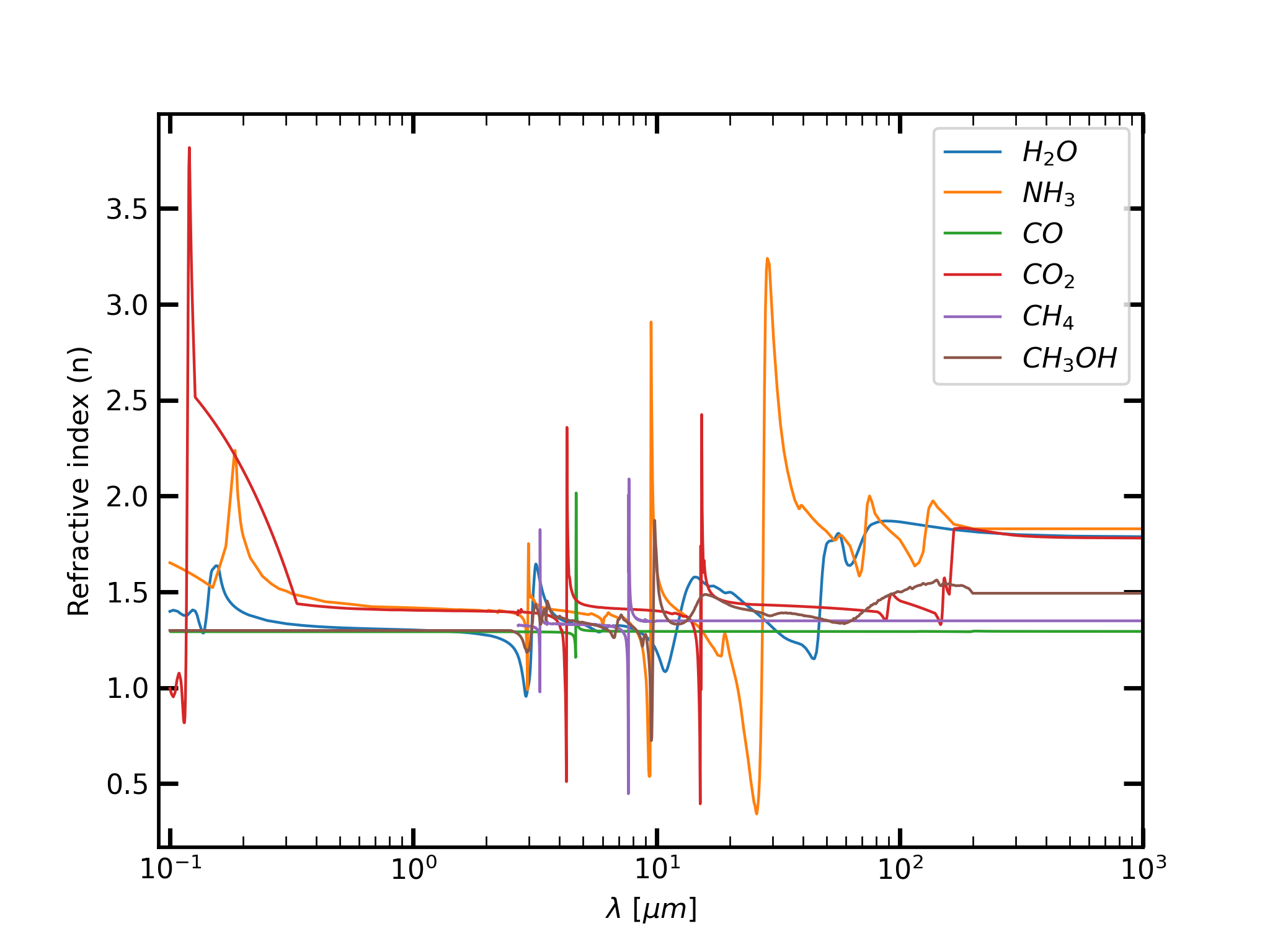}
    \caption{Refractive indices (n) of ices species used in the models.}
    \label{fig:nData}
\end{figure}

\begin{figure}[!ht]
    \centering
    \includegraphics[width=\linewidth, trim=20 5 15 35,clip]{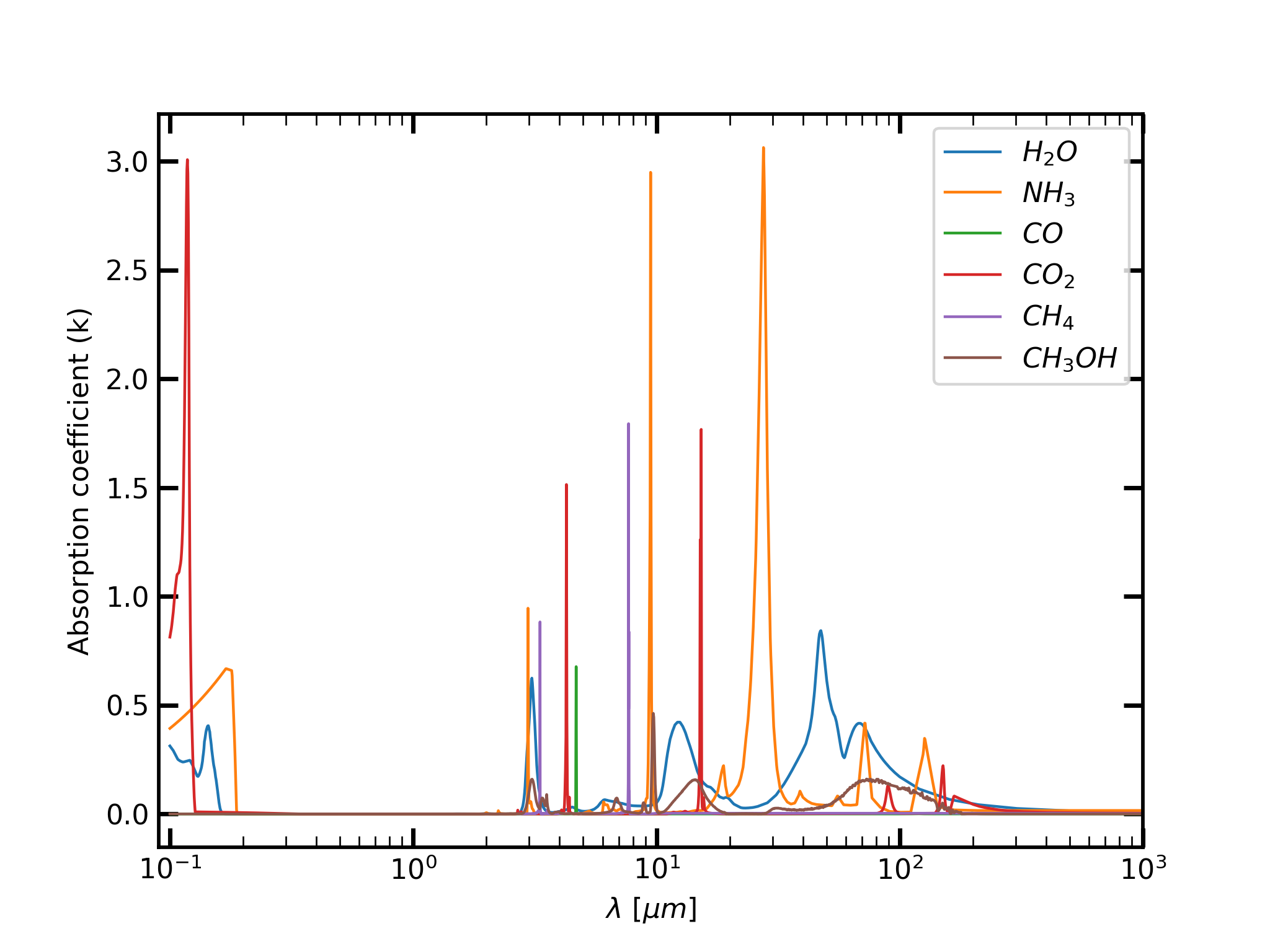}
    \caption{Absorption coefficients (k) of ices species used in the models.}
    \label{fig:kData}
\end{figure}

The source of optical data, the measurement temperature, and structure of six types of ice species and refractory material used are tabulated in Table \ref{tab:iceList}. Ice structures change with temperature, so does the refractive index and absorption coefficient. Though the abundances of the different ices follow the local thermo-chemical state of the disk, the optical constants used are not consistent with temperature. Table~\ref{tab:iceList} summarizes the state and temperature of ices at which the optical data is obtained. It should be noted that the ice optical data do not cover the entire spectrum (0.0912\,$\mu$m - 10$^4$\,$\mu$m). Hence, the data has to be expanded extensively for all species to enable studying all the ice features. For absence of optical data corresponding to a wavelength $\lambda$ which is bounded by optical data at $\lambda_1$ and $\lambda_2$ (such that $\lambda_1<\lambda<\lambda_2$), we perform a log interpolation to estimate the optical data at $\lambda$. In an unbounded case (extrapolation), for longer wavelength side ($\lambda>\lambda_1$) we assume: $n(\lambda)=n(\lambda_1)$ and $k(\lambda)=k(\lambda_1)\cdot\lambda_1/\lambda$; for shorter wavelength side ($\lambda<\lambda_2$) we assume: $n(\lambda)=n(\lambda_2)$ and $k(\lambda)=k(\lambda_2)$. The optical constants used are shown in Fig.\,\ref{fig:nData} and Fig.\,\ref{fig:kData}.

\begin{table}[!ht]
    \centering
    \caption{Optical data of different refractory and ice species.}
    \begin{tabular}{p{0.13\linewidth}p{0.2\linewidth}p{0.15\linewidth}p{0.35\linewidth}}
    \hline
        \textbf{Ice Species} & \textbf{Temperature (K)} &\textbf{ Type }& \textbf{Reference} \\
    \hline
        {$\rm H_2O$} & 266 & Crystalline & \citet{warren1984optical} \\
        {$\rm NH_3$} & 77-88 & Crystalline & \citet{martonchik1984optical}\\
        {$\rm CO_2$} & Multiple & Crystalline & \citet{warren1986optical}\\
        {$\rm CO$} & 10 & Amorphous & \citet{hudgins1993mid,rocha2014determination}\\
        {$\rm CH_3OH$} & 10 & Amorphous & \citet{hudgins1993mid}\\
        {$\rm CH_4$} & 10 & Amorphous & \citet{hudgins1993mid}\\
        \hline
        Carbon & & Amorphous & \citet{zubko1996optical} \\
        Silicate (Mg$_{0.7}$Fe$_{0.3}$SiO$_3$) & & Amorphous & \citet{dorschner1995steps} \\
        \hline
    \end{tabular}
    \label{tab:iceList}
\end{table}

\section{Disk model}
\label{app:typicalTT}
We present a fiducial T Tauri disk ProDiMo model. Table \ref{tab:propertiesTable} presents the disk properties considered for all the models presented in this paper. For more details regarding the parameters, meanings and their values, refer \citet{woitke2009radiation} and \citet{woitke2016consistent}. 

\begin{figure*}[!h]
    \centering
    \includegraphics[width=\linewidth]{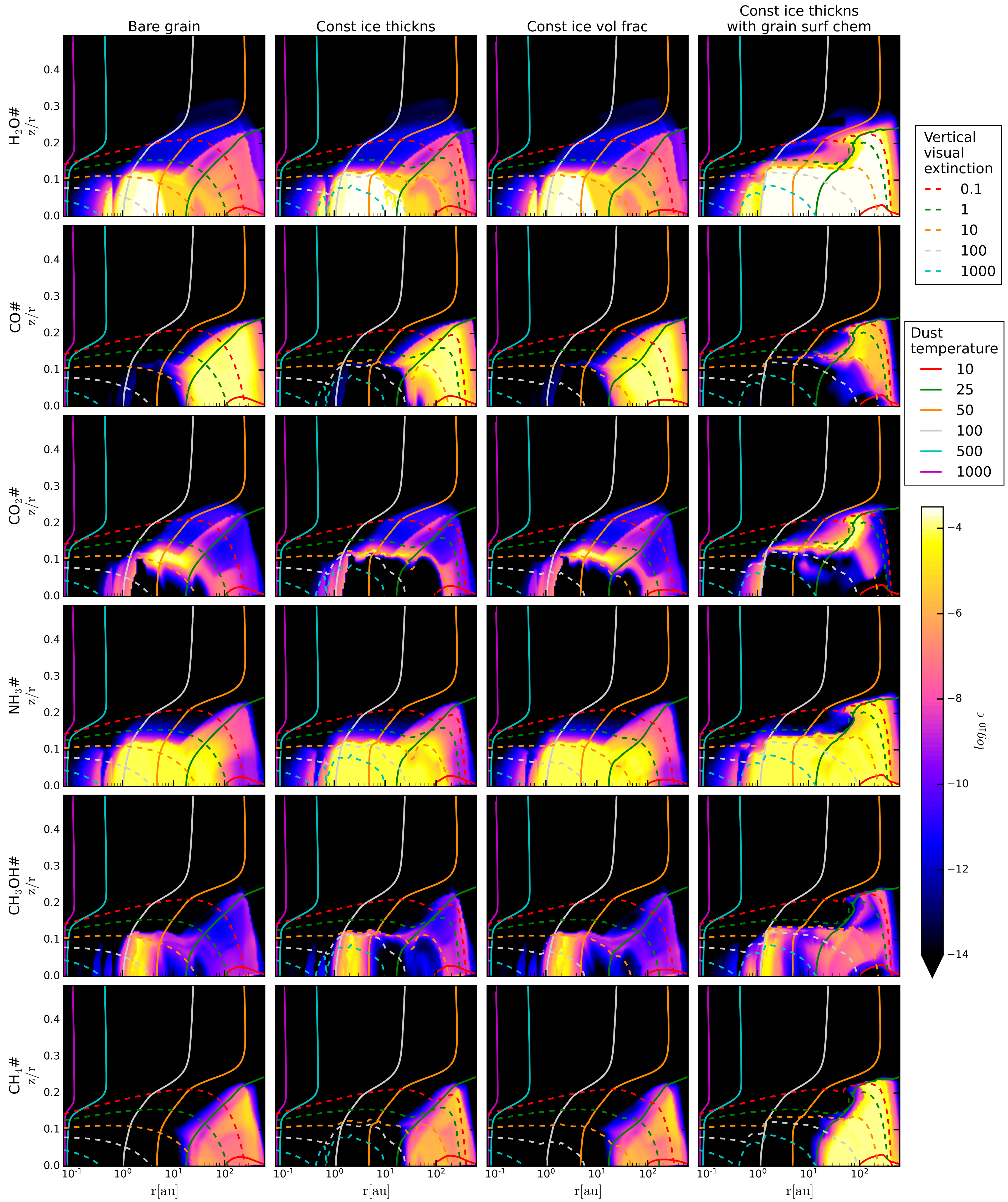}
    \caption{Distribution of the six ice species ($\rm H_2O$, $\rm CO$, $\rm CO_2$, $\rm NH_3$, $\rm CH_3OH$, and $\rm CH_4$) in the presented four models with different ice opacity ($q$) and chemistry settings (See Sect.~\ref{sec:impact}). Each row corresponds to the different ice species and each column corresponds to the one of the four models. The dust temperature and vertical visual extinction contours are also shown with \textit{solid} and \textit{dashed} lines respectively.}
    \label{fig:iceDistIceOpac}
\end{figure*}

\begin{table}
\caption{Common disk parameters for the models.}
\centering
\begin{tabular}{p{0.5\linewidth}p{0.20\linewidth}p{0.15\linewidth}}
\hline \hline
\textbf{Property }             & \textbf{Symbol } & \textbf{Value}\\
\hline
Stellar Mass          & M$_{*}$ & 0.7 M$_{\odot}$ \\
Effective temperature & T$_{*}$ & 4000 K  \\
Stellar luminosity    & L$_{*}$ & 1 L$_{\odot}$ \\ 
UV excess      &    f$_{\rm UV}$     & 0.01           \\
UV powerlaw index      &    p$_{\rm UV}$     & 1.3     \\
X-ray luminosity  & L$_x$ & 10$^{30}$ erg/s \\
X-ray emission temperature & T$_{\rm x}$ & 2$\times$10$^7$ K \\ \hline 
Strength of interstellar UV  & $\chi^{\rm ISM}$ & 1 \\
Strength of interstellar IR  & $\chi^{\rm ISM}_{\rm IR}$ & 0  \\
 Cosmic ray H$_2$ ionization rate & $\zeta_{\rm CR}$ & 1.7$\times$10$^{-17}$ s$^{-1}$  \\ \hline
 Disk Mass & M$_{\rm disk}$ & 0.01 M$_{\odot}$ \\
 Dust/gas mass ratio  & $\delta$ & 0.01 \\
 Inner disk radius  & R$_{\rm in}$ & 0.07 au \\
    Tapering-off radius & R$_{\rm tap}$ & 100 au \\
 Column density power index  & $\epsilon$ & 1 \\
 Reference scale height  & H$_{\rm g}$(100 au) & 10 au \\  
Flaring power index   & $\beta$ & 1.15 \\  \hline
Minimum dust particle radius$\rm ^a$   & a$_{\rm min}$ & 0.05 $\mu$m \\ 
Maximum dust particle radius$\rm ^a$   & a$_{\rm max}$ & 3000 $\mu$m \\ 
Dust size dist. power index$\rm ^a$   & a$_{\rm pow}$ & 3.5 \\ 
Turbulent mixing parameter   & $\alpha_{\rm settle}$ & 0.01 \\ 
Refractory dust composition & Mg$_{0.7}$Fe$_{0.3}$SiO$_3$ & 60\% \\
 & amorph. C & 15\% \\
 & porosity & 25\% \\ \hline
PAH abundance rel. to ISM & f$_{\rm PAH}$ & 0.01 \\
Chemical heating efficiency & $\gamma^{\rm chem}$ & 0.2 \\ \hline
Distance to the observer & d & 140 pc \\ 
\hline
\end{tabular}
\noindent $^{\mathbf{(a)}}$ Parameters related to refractory dust grains.\\
\label{tab:propertiesTable}
\end{table}

\begin{table}
    \centering
    \caption{Adsorption energies of different ice species.}
    \begin{tabular}{p{0.2\linewidth}p{0.2\linewidth}p{0.4\linewidth}}
    \hline
        \textbf{Ice species} & \textbf{Adsorption energy (K)} & \textbf{Reference}\\
    \hline
        {$\rm H_2O$}  & 4800.0 & \citet{2007MNRAS.374.1006B}\\
        {$\rm NH_3$}  & 5534.0 & \citet{2004MNRAS.354.1133C}\\
        {$\rm CO_2$}  & 2990.0 & \citet{2010PhDT........78E}\\
        {$\rm CO$} & 1150.0 & \citet{2004MNRAS.354.1133C}\\
        {$\rm CH_3OH$} & 4930.0 & \citet{2007MNRAS.374.1006B}\\
        {$\rm CH_4$} & 1090.0 & \citet{2010PCCP...12.3164H}\\
        \hline
    \end{tabular}
    \label{tab:adsEnergyList}
\end{table}

\begin{figure}
    \centering
    \includegraphics[width=0.95\linewidth,trim=10 5 0 0,clip]{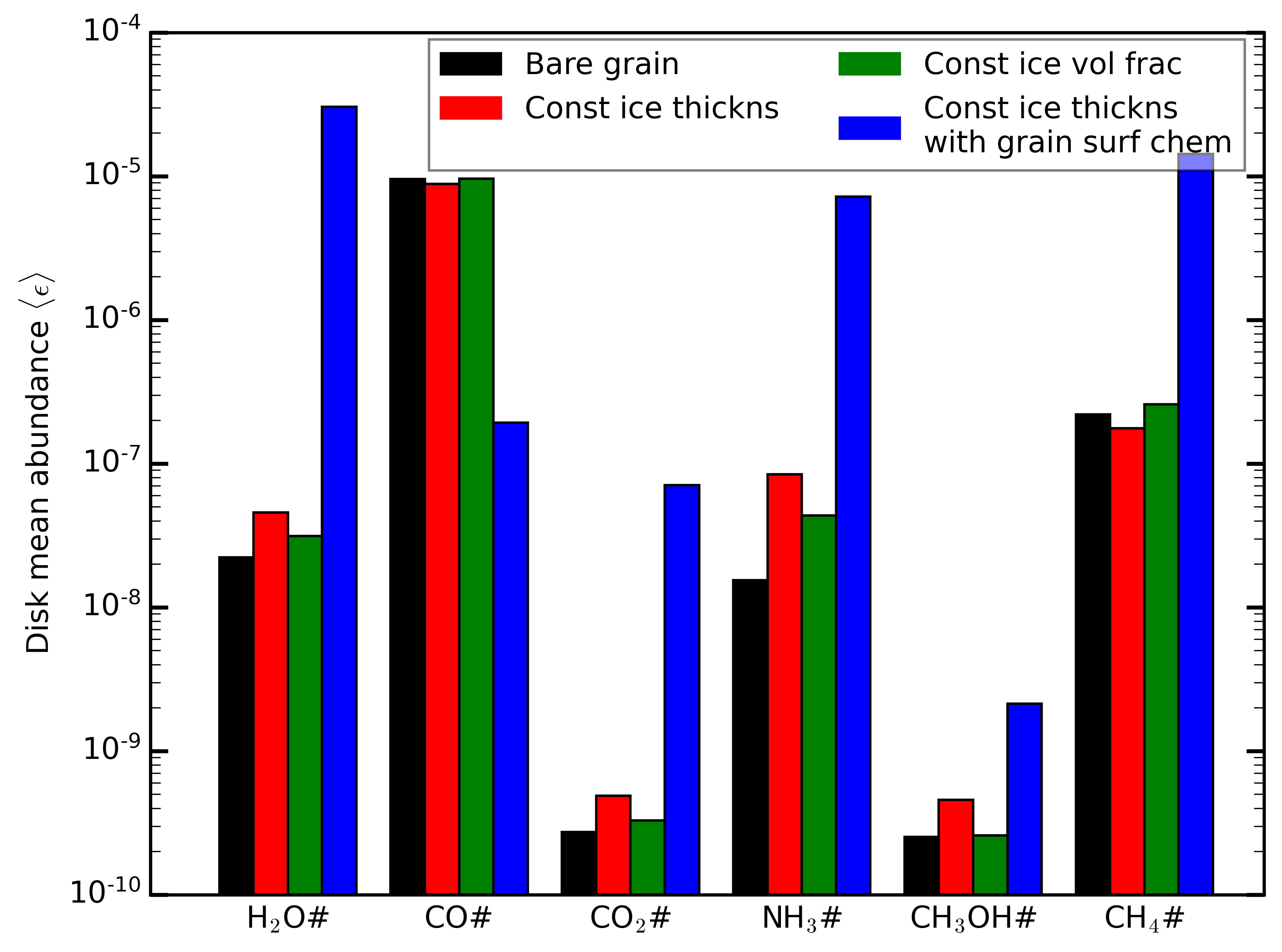}
    \caption{Mean disk abundance of ice species in the four models.}
    \label{fig:meanAbundance}
\end{figure}

\begin{figure}
    \centering
    \includegraphics[width=0.95\linewidth]{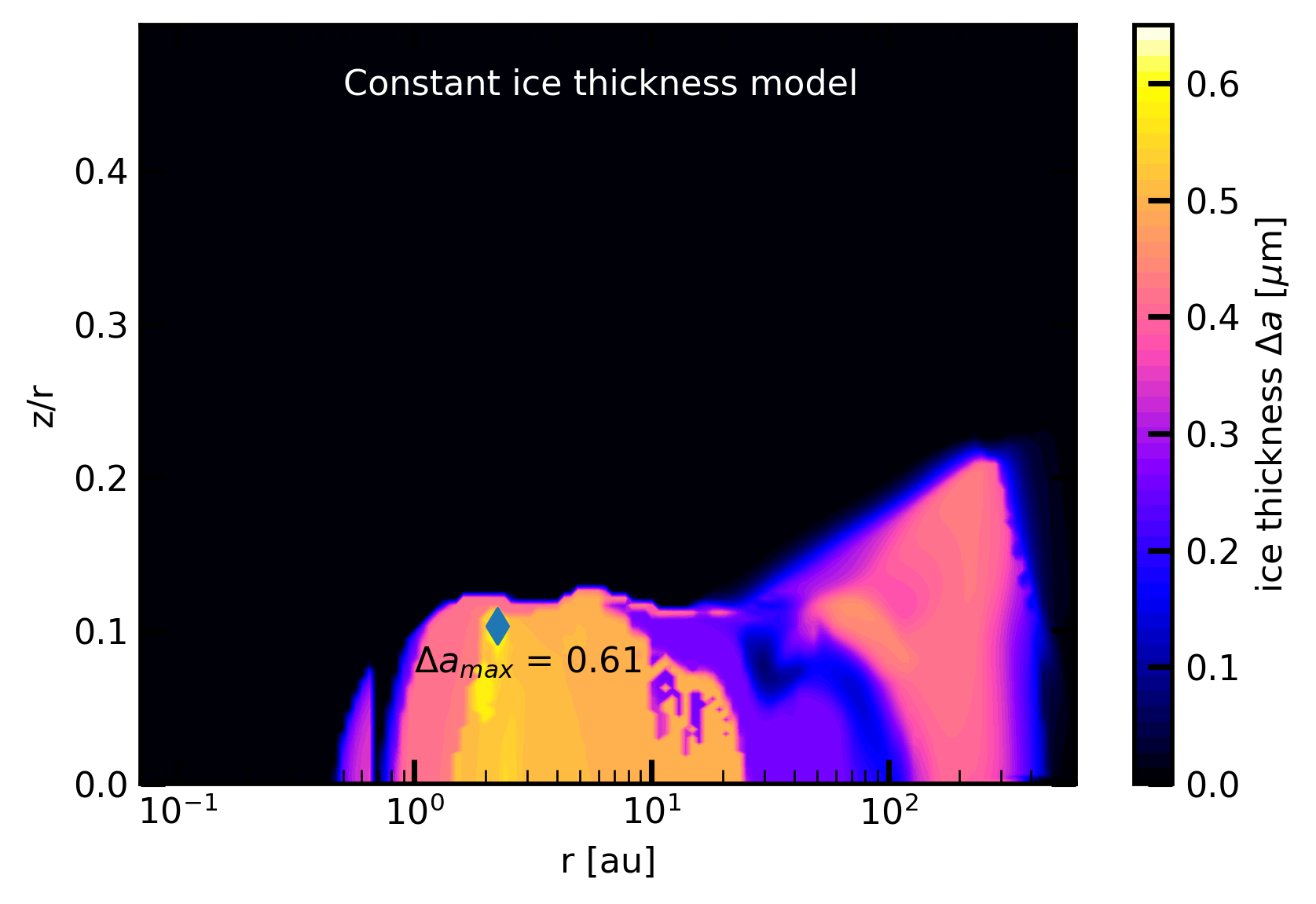}
    \caption{Ice thickness distribution in the constant ice thickness model without grain surface chemistry. The green diamond indicates the location of largest ice thickness (0.61~$\mu$m)}
    \label{fig:iceThickDist}
\end{figure}

\begin{figure}
    \centering
    \includegraphics[width=0.95\linewidth]{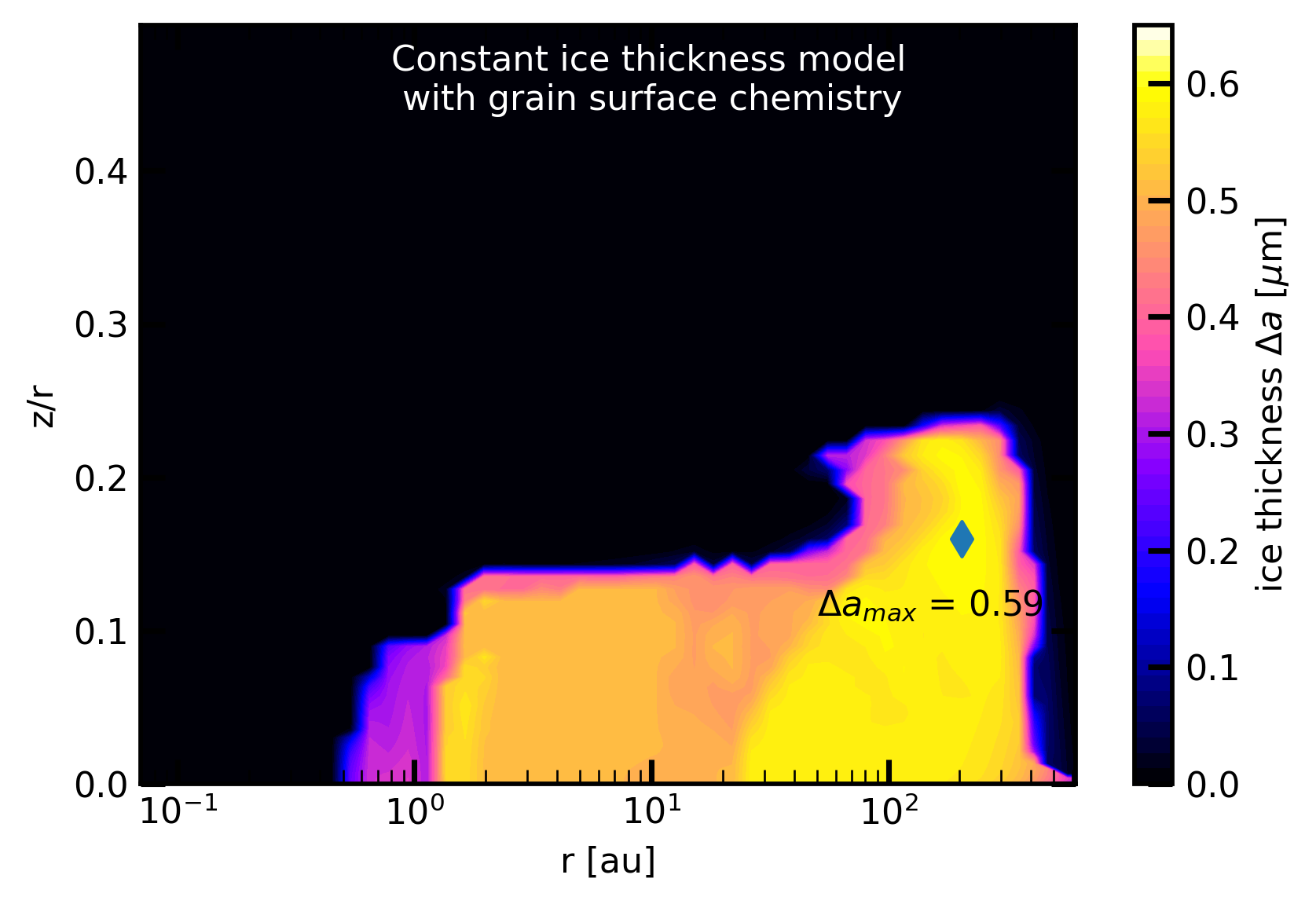}
    \caption{Ice thickness distribution in the constant ice thickness model with grain surface chemistry. The green diamond indicates the location of largest ice thickness (0.59~$\mu$m)}
    \label{fig:iceThickDistGrainSurf}
\end{figure}

\begin{figure}
    \centering
    \includegraphics[width=0.95\linewidth]{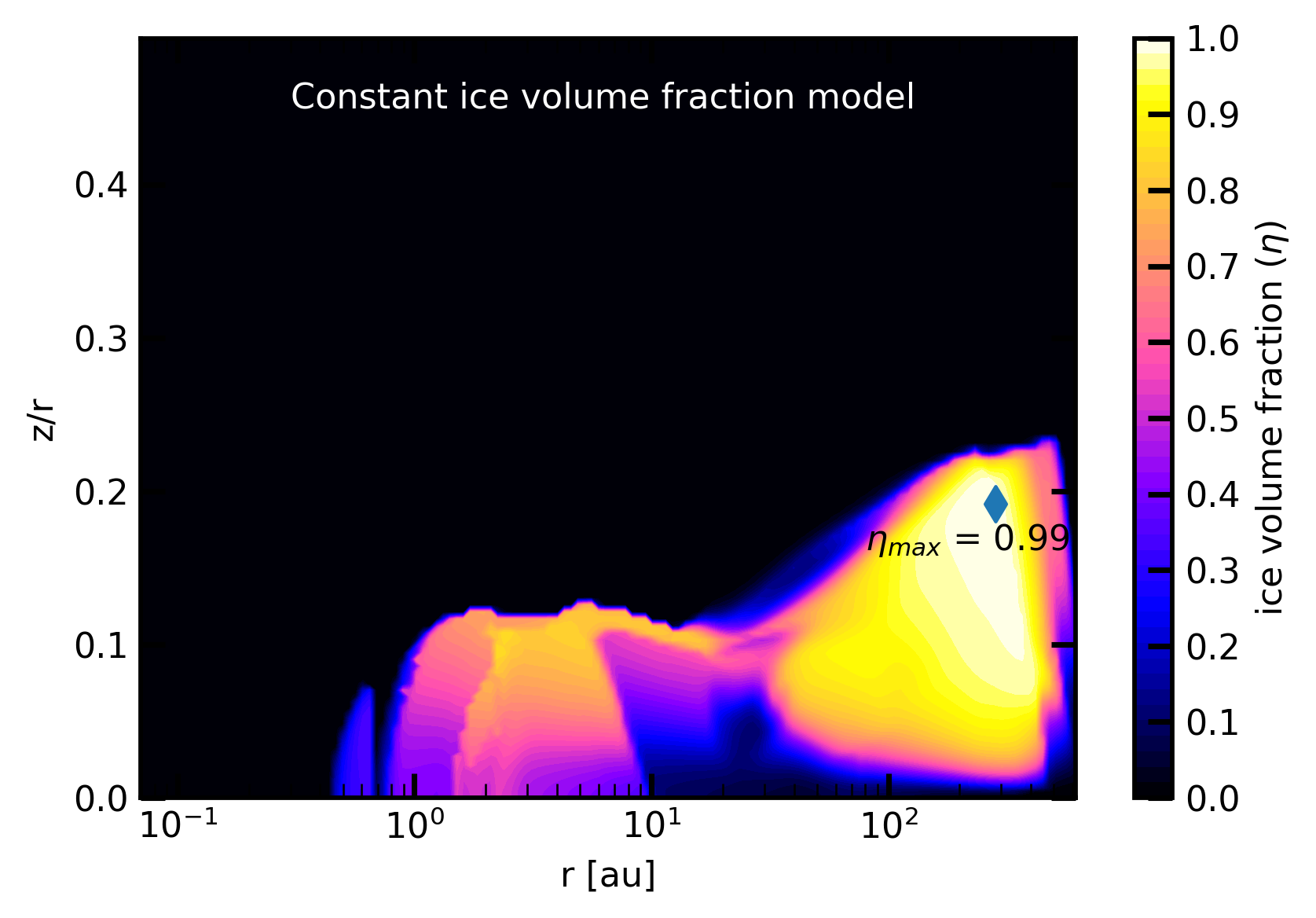}
    \caption{Ice volume fraction distribution in the constant ice volume fraction model. The green diamond indicates the location of largest ice volume fraction (0.99)}
    \label{fig:iceVolFracDist}
\end{figure}

We use the large DIANA chemical network composed of 235 species with 13 elements. The network is described in detail in \citet{kamp2017consistent}. The adsorption energies of different ices considered in this paper are given in Table \ref{tab:adsEnergyList}. The grain surface chemistry used in our models is explained in \citep{2020A&A...634A..42T}, we add two species on top of the large DIANA network: {$\rm H\#$} and {$\rm H_2\#$}, allowing for a minimal surface chemistry, e.g, hydrogenation on surfaces. These models only use physisorption and no chemisorption for the surface chemistry. The purpose of including the surface chemistry is to assess the impact of different chemical networks, and is not to explain the ice observations and the compositions in these objects. We also ran time-dependent chemistry model with grain surface chemistry where the initial composition was determined by a molecular cloud model. We evolved the model up to 2Myr. We noticed that the ice abundances, hence the opacities and the ice features, are very similar to that of the steady state chemistry with grain surface reactions, hence the model is not included in Sect.~\ref{sec:impact}.

Figure \ref{fig:iceDistIceOpac} shows the abundances of {$\rm H_2O$}, {$\rm NH_3$}, {$\rm CO$}, {$\rm CO_2$}, {$\rm CH_3OH$}, and {$\rm CH_4$} ices in the four models discussed in Sect.~\ref{sec:impact}. The figure also shows the dust temperatures and vertical extinction in the disk models. Each column of panels in the figure shows the abundance of the ice species in each model. Models with ice opacities show an increase in water ice in the midplane which extends radially. The outer radii of water ice region extends from 7~au in bare grain model to 25~au, 10~au, and 400~au in models with constant ice thickness, constant ice volume fraction, and constant ice thickness with grain surface chemistry respectively. Fig.~\ref{fig:meanAbundance} shows the mean abundances of the ices in the models presented in this paper. The models with ice opacities are generally more abundant in ices, with exception of {$\rm CO$} and {$\rm CH_4$} ices. The ice abundances are slightly higher in the model with constant ice thickness than the model with constant ice volume fraction by about a factor 1.5. The model with grain surface chemistry has several orders of magnitude higher mean ice abundances, except for {$\rm CO$} ice which is about two orders lower. In general, including ice opacities results in higher ice abundances except for {$\rm CO$} ice. It should be noted that these are mean abundances of the entire disk, hence, essentially are model properties and should not be mistaken for the ice abundances that can be obtained from observations. Figures~\ref{fig:iceThickDist} and \ref{fig:iceThickDistGrainSurf} show the distribution of ice thickness in the disk for the models with constant ice thickness without and with grain surface chemistry respectively. The mean ice thickness for entire the disk (which includes regions without any ice) in these models are 0.045~$\mu$m and 0.094~$\mu$m respectively. Figure~\ref{fig:iceVolFracDist} shows the distribution of ice volume fraction across the disk for the constant ice volume fraction model. The mean ice volume fraction for the entire disk in this model is 24.2\%. Unlike constant ice thickness models, the constant ice volume fraction model shows lower ice volume fraction in the midplane and larger ice volume fraction in the upper midplane. Dust settling leads to an increase in the large grain population in the midplane. Since ice largely resides on large grains in this model, the regions with little large grains yet having significant ice abundances have larger ice volume fractions.

\end{appendix}

\end{document}